\documentclass[a4paper]{article}
\usepackage{graphicx}
\usepackage{amsfonts,amssymb,enumerate,epsf}
\usepackage{authblk}
\usepackage{float}
 \usepackage[english]{babel}
\DeclareMathSizes{11}{9}{7}{5}

 \def\botcaption#1#2{\medskip\centerline{{\scshape #1.}\kern8pt
 {\rm #2}}\bigskip}
\def \ba {\begin{array}}
\def \ea {\end{array}}
\def \be {\begin{equation}}
\def \ee {\end{equation}}

\def\bal{\begin{aligned}}
\def\eal{\end{aligned}}

 \def \R {{\mathbb R}}
  
 \def \Z {{\mathbb Z}}

\def \bk{{\mathbf k}}
\def \bx{{\mathbf x}}
\def \bu{{\mathbf u}}
\def \bh{{\mathbf h}}
\def \bv{{\mathbf v}}
\def \bY{{\mathbf Y}}

 \def \cC {{\cal C}}

 \def \cL {{\cal L}}

 \def \cO {{\cal O}}

\def \La {{\Lambda}}

 \def \p {{'}} 

\def \un {{(1)}}

\def \zer {{ (0) }}
 \def \vet#1 {{\bf {#1}}}

\begin{document}
 
\title{COMPUTER SIMULATIONS FOR THE BLOW-UP OF COMPLEX SOLUTIONS OF THE 3-d NAVIER-STOKES EQUATIONS } 

   \author[*]{C. Boldrighini}
\author[**]{S. Frigio}
\author[***]{P. Maponi}
 \affil[*]{
 Istituto Nazionale di Alta Matematica (INdAM), Gruppo Nazionale per la Fisica Matematica (GNFM), 
 Universit\`a di Roma ``La Sapienza'',
 Piazzale Aldo 2 Moro, 00185 Rome,
 Italy. }
 \affil[**]{Scuola di Scienze e Tecnologie, Universit\`a di Camerino (Italy). Partially supported by COFIN-MIUR}
 \affil[***]{Scuola di Scienze e Tecnologie, Universit\`a di Camerino (Italy)}
\maketitle

 \begin{abstract}
 We present a study by computer simulations of a class of complex-valued solutions  of the three-dimensional Navier-Stokes equations in the whole space $\R^{3}$, which,  according to Li and Sinai \cite {LiSi08},  present a blow-up (singularity) at a finite time.  The computer results allow a detailed study of the blow-up mechanism, and show  interesting features of the behavior of the solutions near the blow-up time, such as the concentration of energy and enstrophy in a small region around a few points of physical space, while outside this region the ``fluid'' remains ``quiet''.  \par \medskip
  
\end{abstract}

 \section {Introduction}
 \label {S1}
 The present paper reports the results of computer simulations for a class of complex-valued solutions  of the   
   incompressible Navier-Stokes (NS) equations in the whole space $\R^{3}$ (no boundary conditions) in absence  of external forces  
  \begin{equation} \label {1}  {\partial {\bf u}\over \partial t} + \sum_{j=1}^3 u_j{\partial\over \partial x_j} {\bf u} = \Delta \mathbf u - \nabla p, \qquad  {\bf x} = (x_1, x_2, x_3)\in \R^3 .  \end{equation} 
  $$ \nabla \cdot \mathbf u = 0, \qquad  \mathbf u(\cdot, 0) = \mathbf u_0. $$
   ${\mathbf u}  =(u_{1}, u_{2}, u_{3}): \R^3 \times [0, \infty) \to \R^3 $ is the velocity field,  $p$ is the pressure,   $\mathbf u_0$ is the initial data, and we assume for the viscosity $\nu=1$ (which is always possible by rescaling). \par\smallskip
   The solutions that we consider were introduced by Li and Sinai \cite{LiSi08}, and their main property is that they become singular at a finite time (blow-up). We believe that their behavior can shed light on a class of related real-valued solutions which share some basic features with their complex analogues. The study of such solutions is under way.   \par\smallskip

 The modern mathematical theory of the Navier-Stokes equations   begins  with a paper of Jean Leray  in 1934 \cite{Leray}.
   Since then, one of the main open questions is whether the solution of the initial value problem in    $\R^{3}$, for smooth initial data and in absence of external forces,  can become singular at a finite time.     
  This
is the celebrated {\it global regularity problem}, which is also one of the seven Millennium Open Problems of the  Clay
Mathematical Institute.
 \par\smallskip
 
   Leray believed
         that  singular solutions with smooth initial data of the Navier-Stokes do exist,  and  are related to  turbulence. Although modern ideas on turbulence developed independently of the problem of singularities, it is clear  
        that  the singular solutions, if they exist,  could be of great importance in the description of fluid motion. Their behavior near the critical time could provide a deeper understanding of physical phenomena  such as a sudden concentration of energy in a small space region, as it happens in a hurricane.    We know in fact \cite{Seregin12} that a loss of smoothness  implies the divergence of the solution at some point of the physical space.
       There is at present no effective model for such phenomena.  
       \par\smallskip
       The main global quantities for the description of the solutions are the total energy $E(t)$ and the total enstrophy $S(t)$, which is a measure of the intensity of the vorticity field $\omega(\mathbf x,t) = \nabla \times \mathbf u(\mathbf x,t)$,
       \begin{equation} \label{2} E(t) = {1\over 2} \int_{\R^{3}} |\mathbf u(\mathbf x,t)|^{2} d\mathbf x , \qquad S(t) =  \int_{\R^{3}} |\nabla \mathbf u(\mathbf x,t)|^{2} d \mathbf x .\end{equation}  \par\smallskip
         By the law of conservation of energy we have
  \begin{equation}\label{equality} E(t) + \int_{0}^{t} S(\tau) d\tau = E(0), \end{equation}
      so that the total energy decreases.  
      Moreover we know   (see \cite{Temam}) that if the total energy and enstrophy of the initial data   $\mathbf u_{0}$ are finite there is a unique regular solution in a maximal interval $t\in [0, T_{c})$, where $0 < T_{c} \leq + \infty$ depends on $\mathbf u_{0}$, and
   if  the  initial energy and  enstrophy are small enough, then $T_{c}   =    +\infty$ (global regularity).  
   Therefore if the critical time
   $T_{c}$ is finite,  the total enstrophy $S(t)$ is unbounded as $t\uparrow T_{c}$.

 \par \smallskip 
 Much work has been devoted to  the global regularity problem for the NS equations, both theoretical and by computer simulations.  Most results  have been obtained for suitable modifications of the equations (see \cite{TT14} \cite{Ch08} and references therein).   A ``close''  result to the existence of a possible  blow-up was 
recently  obtained by T. Tao \cite{TT14}, who proved a finite-time blow-up  for modified NS equations which satisfy the energy conservation. The model is obtained by replacing the quadratic term $\mathbf u \cdot \nabla \mathbf u $   by a suitable average, and is related to the so-called ``dyadic'' model of Katz and Pavlovic, for which a finite-time blow-up can also be proved \cite{KP02}.
\par\smallskip
The  solutions found by Li and Sinai in \cite{LiSi08} have, as we show in detail below, a fairly simple structure in Fourier $\bk$-space.  It is therefore natural to perform the computer simulations in $\bk$-space, and, in particular, it is of great help  the fact that the support of the solutions is concentrated in a thin cone along a fixed direction. Moreover the proofs  in \cite{LiSi08},  which are based on the renormalization group method, give an excellent guideline  for understanding the  ``blow-up mechanism''.   
\par  \smallskip
Complex-valued solutions which blow up in a finite time with a similar behavior in $k$-space have been found also for the Burgers equations \cite{LiSi10} and other models \cite{LiSi10(2)}. Computer simulations of the two-dimensional  Burgers equations have been reported in \cite{BFM12}.\par\smallskip

 The analysis of the Li-Sinai 
 complex solutions suggested the study of a class of real-valued solutions of the NS equations,  which is  in progress. They have the same property of concentration of the support in a thin cone in $k$ space, and, although the question   of possible blow-up's remains open, they show a remarkable growth of the enstrophy for some time interval.  \par\smallskip
 
 We recall that, due to  energy conservation,  for a real blow-up the transfer of energy to the high $k$ modes should be such that the energy remains bounded while the enstrophy increases. 
 For  complex solutions the energy equality (\ref{equality}) holds, but it is not coercive, and in fact the total energy diverges for the Li-Sinai solutions.
 \par\smallskip
 The main difficulties in following the blow-up by computer simulations in $\bk$-space are  due to the fact that,   as we approach the critical time, the support of the solution moves out to infinity. Moreover
  the blow-up is very fast:  it takes place  in a   time of the order of $10^{-5}$ time units. 
The first  computer simulations of a blowup of  the 3d-complex-valued NS equations were carried out by Arnol'd and Khokhlov \cite{AKh09}.
    However, due to computational limitations, their results give only a qualitative description of the blow-up.  It was not possible to give good estimates for the critical time and for the behavior of the energy and the enstrophy near that time.  \par\smallskip
    The plan of the paper is as follows. In \S 2 we formulate the Navier-Stokes equations as an integral equation in $\bk$-space and report  the main features of the Li-Sinai theory on complex blow-up. In \S 3 we describe how the blow-up is detected by computer simulations, with an estimate of the critical time. \S 4 and \S 5 are devoted to the behavior of the solutions near the critical time in $\bk$- and in $\bx$-space, respectively.
    \S 6 gives some technical details on the computation, and \S 7 is devoted to  concluding remarks.

 \section {The Li-Sinai predictions for complex solutions}
 \label {S2}
 
As a guideline to the understanding of the main features of the solutions, we briefly describe the Li-Sinai theory. We refer the reader to the paper \cite{LiSi08} for the proofs and further details.
\par\smallskip

      We write the NS equations  in terms of the transform
      \be\label{vu} \mathbf v(\mathbf k, t) = {i\over (2\pi)^{3}} \int_{\R^3} \mathbf u(\mathbf x, t) e^{i \langle \mathbf k, \mathbf  x\rangle } d{\mathbf x}, \qquad \mathbf k= (k_1, k_2, k_{3}) \in \R^3\ee
    where $\langle \cdot, \cdot \rangle$ denotes the scalar product in $\R^3$. The normalization is chosen in such a way that the transform of a product of functions is the convolution in $\bk$ of the transforms, with no extra factor.\par\smallskip
    
         The NS equations (\ref{1})   go,  by means of a Duhamel formula, into the following integral equation 
   $$ \mathbf v ({\bf k},t) = e^{-t {\bf k}^2} \mathbf v_{0}({\bk})   + $$
     \be\label{kequation} +  \int_0^t e^{-(t-s) {\mathbf k}^2} ds \int_{\R^3}\langle \mathbf v(\bk-\bk\p, s),\mathbf k \rangle    P_{\mathbf k}   \bv(\bk\p, s)   d \bk\p , \ee
    where the initial condition $ \bv_{0}(\bk)$ is the transform of $\bu_{0}$, and $P_{\bk}$  is the solenoidal projector  expressing incompressibility     $$P_{\mathbf k} \mathbf v = \mathbf v - {\langle \mathbf v, \mathbf k\rangle \over {\bf k}^2} \mathbf k . $$
        
      \par\smallskip
   We consider real solutions of the equation (\ref{kequation}), which correspond to  complex solutions $\bu(\bx,t)$. 
The initial data   $\bv_{0}$ are  localized around a point $\bk^{\zer}$, at a certain distance from the origin.  For our computer simulations  we always  took $\bk^{\zer}= (0, 0, a)$ with  $5\leq a  \leq 25$, and the  support of $\mathbf v_{0}$  in a circle with center $\bk^{\zer}$ and radius $r<a$.   \par\smallskip
 Multiplying the initial data  $\bv_{0}$ by  a positive parameter   $A$ and iterating the Duhamel formula we can write the solution  as a power series 
 \begin{equation}\label{serie} \bv_{A}(\bk, t) =  A   e^{- t {\bk}^2} \bv_{0}(\bk) + \int_0^t e^{-{\mathbf k}^2(t-s) } \sum_{p=2}^\infty A^p \mathbf g^{(p)}(\mathbf k, s) ds. \end{equation}
     Substituting into the equation,   we see that  the functions   $\mathbf g^{(p)}(\mathbf k, s)$ satisfy a recursive relation of convolution type. Setting $\mathbf g^\un(\mathbf k, s) = e^{- s {\bk}^2} \bv_{0}(\bk)$      and
 $$  \mathbf g^{(2)}(\mathbf k, s) =   \int_{\R^3} \left \langle \bv_{0}(\bk-\bk\p),\bk \right \rangle   
  P_{\bk} \bv_{0}(\bk\p)   e^{- s(\bk - \bk\p)^2 - s (\bk\p)^2} d {\bk}\p ,$$
 we find for $p>2$ the recursive relation
 
 $$    \mathbf g^{(p)}(\mathbf k, s) =  $$
 $$ =  \int_0^s ds_2 
    \int_{\R^3}  \left \langle \bv_{0}(k-k\p),\bk \right \rangle    P_{\bk}  \mathbf g^{(p-1)}(\bk^\p, s_{2}) e^{- s(\bk - \bk\p)^2 - (s-s_{2}) {(\bk}^\p)^2} d {\bk}^\p  + $$
 \be\label{0}+  \sum_{p_1 + p_2 = p\atop p_1, p_2 >1} \int_0^s ds_1 \int_0^s ds_2 
   \int_{\R^3}\left  \langle \mathbf g^{(p_1)}(\mathbf k - \mathbf k\p, s_1), \bk \right \rangle \cdot $$ $$\cdot   P_{\bk}\mathbf g^{(p_2)} (\bk\p, s_2) e^{-  (s-s_1) (\bk - \bk\p)^2 - (s-s_2) ( \bk\p)^2} d\mathbf k\p  + \ee
 $$ +   \int_0^s ds_1 \int_{\R^3}  \left  \langle \mathbf g^{(p-1)}(\bk - \bk\p, s_1), \bk \right  \rangle   P_{\bk} \bv_0(\bk\p) e^{-  (s-s_1) (\mathbf k - \mathbf k\p)^2 - s ( \bk\p)^2} d\bk\p.$$

\par\smallskip

 Observe that if $C= \rm{supp} \; \bv_{0}$, then, by iteration of the convolution, the support of $\mathbf g^{(p)}$ will be $\underbrace {C + \ldots + C}_{ p \;  times}$.  As the support of $\bv_{0}$ is around $\bk^{\zer} = (0,0, a)$,   the support of the solution extends along the $k_{3}$-axis.  \par \smallskip 
 
 By analogy with the  theory of probability, where the convolution is the distribution of a sum of random variables, 
 we know that for
 large $p$ the  support of $\mathbf g^{(p)}$ is  around  $p \bk^{\zer}$, in a region with transversal dimensions of the order $\sqrt {p}$. 
Moreover if $p$  is large  the terms of the sum for which $\max \{ p_{1}, p_{2}\} \leq p^{1\over 2}$  can be neglected, and the Gaussian densities   give a significant contribution to the integrals only for $s_{1}, s_{2}$ near the endpoint $s$. Therefore we introduce the  new variables and  functions 
$$\bk = p \bk^{\zer} + \sqrt {p}   \mathbf Y, \qquad \mathbf h^{(p)}(\mathbf Y, s) = \mathbf g^{(p)}( p \bk^{\zer} + \sqrt {p }  \bY, s) $$
$$ s_{j} = s  \left ( 1 - {\theta_{j}\over p^{2}_{j}} \right ), \qquad j=1,2 . $$
 \par \smallskip
 
Integrating over $\theta_{j}, j=1,2$ and  setting $\gamma = {p_{1}\over p}$ we get 
\be\label{4}  \mathbf h^{(p)} (\bY, s) =   {p^{5\over 2}}  \sum_{p_{1}+p_{2}= p\atop p_{1},p_{2} > \sqrt p}   {1\over p_{1}^{2} p_{2}^{2}} \int_{\R^{3}} P_{ \mathbf e_{3}+ {\bY \over \sqrt {p}}}   \mathbf h^{(p_{2})} \left ( {\bY\p\over \sqrt {1-\gamma}}, s \right )\cdot   \ee 
$$\cdot  \left \langle \mathbf h^{(p_{1})}   \left ( {\bY - \bY\p\over \sqrt \gamma}, s \right ), \mathbf e_{3}+ {\bY \over \sqrt {p}} \right \rangle  d\bY\p \; \left (1 + o(1) \right ),$$
where $\mathbf e_{3}= (0, 0, 1)$.
 As  $\bh^{(p)}$ is orthogonal to $\bk = ( \sqrt {p}  Y_{1}, \sqrt {p}  Y_{2}, pa + \sqrt {p}  Y_{3} )$, by incompressibility,  we also set 
 \be\label{5}\bh^{(p)} (\bY, s)= \left ( H^{(p)}_{1}(\bY, s),  H^{(p)}_{2}(\bY, s), {F^{(p)}(\bY, s)\over \sqrt {p} \; a}  \right ),\ee
 and $F^{(p)}(\bY, s)$ is of finite order:
$$ Y_{1} H^{(p)}_{1}(\bY, s) + Y_{2} H^{(p)}_{2}(\bY, s) + F^{(p)}(\bY, s) = \cO(p^{-{1\over 2}}a^{-1}) .$$ 
Therefore 
  $\mathbf h^{(p)} (\bY, s)$,  is essentially transversal to the $k_{3}$-axis, and as $p\to\infty$,
    $P_{ \mathbf e_{3}+ {\bY \over \sqrt {p}}}   \mathbf h^{(p_{2})} \to  \mathbf h^{(p_{2})} $,  i.e., the solenoidal   projector in (\ref{4}) tends to      the identity.   
  \par \smallskip 
   The fundamental {\it Ansatz} is that for some set of initial data $\mathbf v_{0}$, when  $p$ is large and  $s$ in some interval of time,  the recursive relation (\ref{4}) has an approximate  solution which is asymptotically of the form
 \begin{equation}\label{ansatz} \mathbf h^{(p)}(\bY, s) = Z \;    p \; (\Lambda(s))^{p}  \prod_{j=1}^{3}  g_{\sigma_{j}}(Y_{j})  \left ( \mathbf H(\bY) + \mathbf \delta^{(p)}(\bY, s) \right ) .\end{equation}
 Here $Z$ is a suitable constant, $\La(s)$ is a  function of time, which will be discussed below,    $g_{\sigma}(x) ={ e^{- {x^{2}\over 2 \sigma}}\over  \sqrt {2\pi\sigma} } $ denotes the centered Gaussian density on $\R$, $\sigma_{1}, \sigma_{2}, \sigma_{3}$ are positive constants,   $\mathbf H$ is a   vector function independent of time,  orthogonal to   $\mathbf e_{3}$,  and depending only on $Y_{1}, Y_{2}$,
 $$\mathbf H(\bY) = \left (H_{1}(\bY),   H_{2}(\bY), 0 \right ),$$ 
 and the remainder \begin{equation} \label {6}\delta^{(p)}( \bY, s) =
  \left (  \delta_{1}^{(p)}(\bY, s),    \delta_{2}^{(p)}(\bY, s),   \delta_{3}^{(p)}(\bY, s)\right ) \end{equation}
  is    such that $\delta^{(p)}(\bY, s) \to 0$ as $p\to \infty$.  
  \par\smallskip
 Observe that by the  {\it Ansatz} (\ref{ansatz}), the function  $\mathbf h^{(p)}(\bY, s)$ is proportional to a product of Gaussian functions, and the time dependence of its leading term is determined by the  function $\Lambda(s)$.

 \par\smallskip
In view of possible rescalings 
it is not restrictive to set  $\sigma_{i}=1, i=1, 2, 3$.  Inserting (\ref{ansatz}) into (\ref{4}), treating $\gamma$ as a continuous variable, neglecting the remainders, choosing the constant $Z$ in  a suitable way, and integrating over $Y_{3}$,  one can see that  $\mathbf H(\bY)$ is a solution of the  integral fixed point equation
 \be\label{7} g_{1}(\bY)  \mathbf H(\bY) =\int_{0}^{1} d\gamma\int_{\R^{2}} g_{\gamma}(\bY-\bY\p) g_{1-\gamma}(\bY\p)   \cL(\mathbf H; \gamma, \bY, \bY\p)   \mathbf H\left ( {\bY\p\over \sqrt{1-\gamma}}\right ) d\bY\p\ee
 where, by abuse of notation, we write   $\bY = (Y_{1}, Y_{2})$,  $ g_{\sigma}(\bY) = {e^{- {Y_{1}^{2}+Y_{2}^{2}\over 2 \sigma} }\over 2\pi \sigma}$,  and   $$  \cL(\mathbf H; \gamma, \bY, \bY\p)  =  (1-\gamma)^{3\over 2} \left \langle {\bY - \bY\p\over \sqrt \gamma},   \mathbf H \left ({\bY - \bY\p\over \sqrt \gamma} \right ) \right \rangle+$$
 $$+   \gamma^{1\over 2} (1-\gamma) \left \langle { \bY\p\over \sqrt {1-\gamma}},   \mathbf H \left ({\bY\p\over \sqrt {1-\gamma}} \right ) \right \rangle.$$
  \par \smallskip

  The solutions, or ``fixed points'',  of the functional equation (\ref{7})  are found by expanding $\mathbf H$ in Hermite polynomials ${\rm He}_{k}, k=0,1,\ldots$, which are orthogonal with respect to the standard gaussian,
 \be\label{hermite} H_{j}(\bY) = \sum_{m_{1}, m_{2}=0}^{\infty} \ell^{(j)}_{m_{1}m_{2}} {\rm He}_{m_{1}}(Y_{1})  \; {\rm He}_{m_{2}}(Y_{2}), \qquad j=1,2 .\ee
 As shown  in \cite{LiSi08}, there are infinitely many fixed points, and there is a  class $\cC$ of  them such that  for each choice of a fixed point in $\cC$ there is an open set of initial data $\bv_{0}$ for which  the  solution 
satisfies  the {\it Ansatz} (\ref{ansatz}) for $s\in S = [s_{-}, s_{+}]$, where $S$ is a non-empty  time interval. 
The open set is constructed by linearized stability analysis, and the proofs are based  on the renormalization group method.\par\smallskip
  As for the time dependence of the solutions, a delicate analysis  shows that the function $\Lambda(s)$ is differentiable and strictly increasing (see \cite{LiSi08} and references therein).
  Setting $A={1\over \Lambda(\tau)}$, for $\tau \in S$, it can be shown that
  the tail of the series appearing in equation (\ref{serie}),  for $p>p_{0}$, with $p_{0}$ large enough, can be replaced near the critical time $\tau$ by the asymptotics
    \be\label{8}    \sum_{p=p_{0}}^\infty A^p \mathbf g^{(p)}(\mathbf k, s)  \approx  C \;   \sum_{p=p_{0}}^\infty p    \left ( {\La(s)\over \La(\tau)} \right )^{p} g\left ({\bk - p \bk^{(0)}\over \sqrt { p} } \right )  \mathbf H \left ({\bk - p \bk^{(0)}\over \sqrt { p} } \right ), \ee
where $C$ is a constant, $\mathbf H$ is the chosen fixed point,  and $g$   is the three-dimensional standard Gaussian distribution.\par\smallskip
The explicit asymptotics (\ref{8}) shows that the main support of the solution extends along the direction $\mathbf k^{\zer}$, i.e., along the positive $k_{3}$-axis in a thin cone of transversal diameter proportional to $\sqrt {k_{3}}$. \par\smallskip
The solution blows up at the critical time $T_{c}= \tau$. In fact, as $\Lambda(s)$ is differentiable and strictly increasing, we have,  as $s\uparrow \tau$
\be\label{smooth} \ln {\Lambda(s)\over \Lambda(\tau)} = - {\Lambda^{\prime}(\tau)\over \Lambda(\tau)} (\tau-s) (1 + r(\tau-s))\ee
where $r$ is continuous and $r(0)=0$. Therefore if 
$\bk $ is close to $p \bk^{(0)}$  the  factor multiplying the fixed point function $\mathbf H$ is of   order $ e^{-\kappa|\bk|(\tau-s)}|\mathbf k|$, with $\kappa >0$.  This quantity is maximal for $|\bk| \approx {{\rm const}\over \tau-s}$, so that  
  the main support of the function escapes to infinity as $s\uparrow \tau$. 
The solution in $\mathbf x$ space, i.e., the inverse  Fourier transform $\mathbf u(\mathbf x,t)$ of $\mathbf v(\mathbf k, t)$  has its main support near  the origin, 
where its energy, as shown below,  is concentrated in a small region.\par\smallskip
An important observation for what follows is that it can happen for some initial data that the fundamental  {\it Ansatz} (\ref{ansatz}) holds in modified form, in which the term $(\Lambda(s))^{p}$ is replaced by $(-1)^{p} (\Lambda(s))^{p}$, where again $\Lambda$ is a positive increasing function. In fact the recursive relation (\ref{4}) is unchanged if we replace $h^{p}$ with $(-1)^{p} h^{p}$. 
In this case the right side of the asymptotics (\ref{8}) is replaced by the expression
\be\label{8a} C \;   \sum_{p=p_{0}}^\infty (-1)^{p} p    \left ( {\La(s)\over \La(\tau)} \right )^{p} g\left ({\bk - p \bk^{(0)}\over \sqrt { p} } \right )  \mathbf H \left ({\bk - p \bk^{(0)}\over \sqrt { p} } \right ). \ee
Such solutions  also blow up, but their behavior, as we shall show below,  is  different from that of the solutions for which the asymptotic series with positive coefficients (\ref{8}) holds.

     \section {Computer Simulations: detecting the blow-up}
 \label {S3}
   We simulate the integral equation (\ref{kequation}) in Fourier space. As we said above, the support of the solutions in $\bk$-space is concentrated in a thin cone along the $k_{3}$-axis. This fact greatly simplifies the computer simulations. In fact, if we extend the integration region along the direction $\bk^{\zer}$ by a factor $D>1$, we only need to extend it in the transversal direction by a factor a little larger than $\sqrt D$. \par\smallskip
    The discretization in $\bk$-space is implemented by a regular mesh of points containing the origin, and chosen in such a way that it contains the region where the solution is significantly non-zero. 
 The results of the computer simulations appear very stable with respect to refinements of the mesh, in accordance with the fact  that   the solution $\bu(\bx,t)$ in $\bx$-space is essentially concentrated in a small region around the origin (see \S 5).  They are however  sensitive to the time step, which  has to be refined for stability  as we approach  the critical time.   \par \smallskip
 
 For all simulations reported below the mesh in $\bk$-space is taken with nearest neighbor distance $1$, and is a set of the type  $ R = [-127, 127] \times [-127, 127] \times [-19, L] \subset \Z^{3}$ (the brackets $[\ldots ]$ denote  intervals in  $\Z$), with $L= 2028, 2528$.  Control simulations with mesh step $1/2$ and with $L= 3000$ were performed to check stability.   We also checked that the time step $\delta_{t}= 10^{-7}$  is small enough to ensure stability. 
   The results in this range of values of $L$ are  stable  up to times sufficiently  close to the critical time.    \par\smallskip
 
Most of the simulations were done with  initial data $\bv_{0}$  concentrated around the point $\mathbf k^{(0)}=(0, 0, 20)$, with support in the circle $|\bk - \bk^{\zer}|\leq 17$. 
  We only report the results  obtained with the following choices
 \be\label{iniziali} \bv_{0}^{\pm}(\bk) = \pm  C\;  \bar{\bv}_{0}(\bk), \qquad  \bar{\bv}_{0}(\bk) =  \left (- k_{1}, -k_{2}, {k_{1}^{2}+ k_{2}^{2} \over k_{3} }\right ) {e^{-{(\bk - \bk^{\zer})^{2}\over 2}}\over (2\pi)^{3\over 2}  }, \ee 
 for different values of   the positive constant $C$, which controls the initial energy $E_{0}$ and the initial enstrophy. Observe that $\bar \bv_{0}(\bk)$ is proportional to the solenoidal variant of the vector $\mathbf H^{\zer}(\bk-\bk^{\zer})$, obtained by adding the resulting third component. This choice is   the simplest variant of the prescription in \cite{LiSi08} (\S 7, formula (39)), and it gives good results for computer simulations.\par\smallskip
 As we shall see, the initial data $\bv_{0}^{\pm}$ lead to asymptotics of the type (\ref{8a}) and (\ref{8}), respectively. 

    \par\smallskip

 A full screening for the ``best'' cases with a large random choice of the initial parameters  in the admissible region, as in our previous paper on the Burgers equations \cite{BFM12}, was not possible because for the three-dimensional Navier-Stokes  it takes too much computer time.
 \par\smallskip
 
  We  examined about a hundred initial data, chosen according to the prescriptions  in \cite{LiSi08}.  All of them lead to the fixed point  $\mathbf H^{\zer}= -2 (Y_{1}, Y_{2},0)$, corresponding to the following  choice of the parameters in the expansion (\ref{hermite}):  $\ell^{1}_{10}= \ell^{2}_{10}= -2$,  and all the other components $\ell^{(j)}_{m_{1}m_{2}}$ are set to zero.

  \par\smallskip
In describing the behavior of the solutions near the blow-up  an important role is played by  the total energy and the total enstrophy (\ref{2}), and, in view of the structure of the solutions,  by the marginal energy and enstrophy densities along the third axis  in $\bk$-space
 \be\label{3k} E_{3}(k_{3}, t) = \int_{\R\times \R} d k_{1} dk_{2} e(\mathbf k, t) \qquad S_{3}(k_{3}, t) = \int_{\R\times \R} d k_{1} dk_{2} s(\mathbf k, t) ,\ee
$$ e(\bk, t) = {1\over 2} |\bv(\bk, t)|^{2}, \qquad s(\bk, t) =  |\bk|^{2} |\bv(\bk, t)|^{2}.$$
 Observe that by the definition of the transform (\ref{vu}) the total energy and enstrophy in (\ref{2}) are given by
 $$E(t) = {(2\pi)^{3}\over 2} \int_{\R^{3}} e(\mathbf k, t) d\bk , \qquad S(t) = (2\pi)^{3} \int_{\R^{3}} s(\mathbf k, t) d\bk.$$
The corresponding quantities  in $\bx$-space are
\be\label {3x}\tilde E_{3}(x_{3}, t) = \int_{\R\times \R} d x_{1} dx_{2} \tilde e(\mathbf x, t) \qquad \tilde S_{3}(x_{3}, t) = \int_{\R\times \R} d x_{1} dx_{2} \tilde s(\mathbf x, t) ,\ee
$$ \tilde e(\bx, t) = {1\over 2} |\bu(\bx, t)|^{2}, \qquad \tilde s(\bx, t) =  \left |\nabla \bu(\bx, t)\right |^{2}.$$
 We  also consider the transverse marginals $E_{j}(k_{j},t)$, $\tilde E_{j}(x_{j},t)$, $S_{j}(k_{j},t)$, $\tilde S_{j}(x_{j},t)$, $j=1,2$. which are defined in an obvious way.
\par\smallskip
 
For the initial data (\ref{iniziali}),  if the constant $C$ is large enough, corresponding for the initial energy to $E_{0} > 2500$,   the solution  blows up after a time of the order $10^{-3}-10^{-4}$ time units.   
 Observe however  that in our screening  we never went beyond a time of the order $10^{-2}$, so that it may well be that some of the cases with low initial energy do in fact blow up at a later time.

 \par \smallskip
The behavior of the solution  coming out of the initial data   does not change qualitatively when we increase the constant $C$ in (\ref{iniziali}) beyond the critical value. Therefore we only report results   for   initial energy $E_{0}= 200 \times (2\pi)^{3}\approx 496 \times 10^{3}$.\par\smallskip

\par\smallskip

For an estimate of the critical time  $\tau$ we looked first at the behavior in time of the total energy and the total enstrophy. Fig. 1 and Fig. 2 give the behavior of those quantities on a logarithmic scale, for the initial data $\bv_{0}^{+}$ and $\bv_{0}^{-}$, respectively.  Observe that the enstrophy starts growing significantly earlier than the energy.

\par\smallskip
As we see, for the initial data $\bv_{0}^{-}$ the fast growth starts earlier, at about $t\approx 10 \times 10^{-5}$,  than  for $\bv_{0}^{+}$, for which it starts at about  $t\approx 15 \times 10^{-5}$. Also, for  $\bv_{0}^{-}$ the growth is faster. This is to be expected, because, as shown in the next paragraph, for the solution with initial data $\bv_{0}^{+}$ the asymptotics with alternate signs  (\ref{8a}) holds.  In what follows, we mostly report results of simulations with initial data $\bv_{0}^{+}$, which are somehow easier to follow. 

  \par\smallskip 
 The  growth of the total energy and total enstrophy near the critical time is predicted in   \cite{LiSi08}  to be  as a power law, i.e.,  
  \be\label{prediction}   E(t) \sim {C_E \over (\tau - t)^5}, \qquad  S(t) \sim {C_S \over (\tau - t)^7},\ee 
  where $\tau$ is the critical time and $C_{E}, C_{S}$ are  positive constants.\par\smallskip
   It is however unclear from the analysis in \cite{LiSi08} whether the asymptotics (\ref{prediction}) holds with the same powers for initial data of the type $\bv_{0}^{+}$, which give rise to the series with alternate signs (\ref{8a}). It is not easy to determine the exact powers by  numerical analysis, and we need be content with compatibility estimates.

 The regression plots for   $(E(t))^{-{1\over 5}}$ and  $(S(t))^{-{1\over 7}}$   vs $t$ are shown  in Figg. 3, 4, for $\bv_{0}^{+}$ and in Figg. 5,6 for $\bv_{0}^{-}$.  The regression is restricted to the range of times with approximately linear behavior. As we shall see in the next section, as $t$ gets close to the critical time a significant amount of ``mass'' gets out of the integration region, so that the growth seen in the simulations slows down.    This is also the reason why, as shown in Figg. 3,4, the estimates of $\tau$, given by the intercept with the horizontal axis, decreases with $L$.  As the estimates for the two values of $L$ are close to each other, for the initial data $\bv_{0}^{+}$ we may take as un upper estimate for the critical time the value $\tau_{+} =  1726  \times  10^{-7}$. 
 \par\smallskip
 The $R^{2}$ parameter in  the linear regressions  shown in Figg. 3-6 is  always around $0.999$, so that we may say that  the computer results are compatible with the predictions (\ref{prediction}). \par\smallskip
   
   In the next paragraph we give other estimates of the critical time based on a  different method.
\par\smallskip

 \begin{figure}[H]  
\centerline {
\includegraphics[width=3.5in]{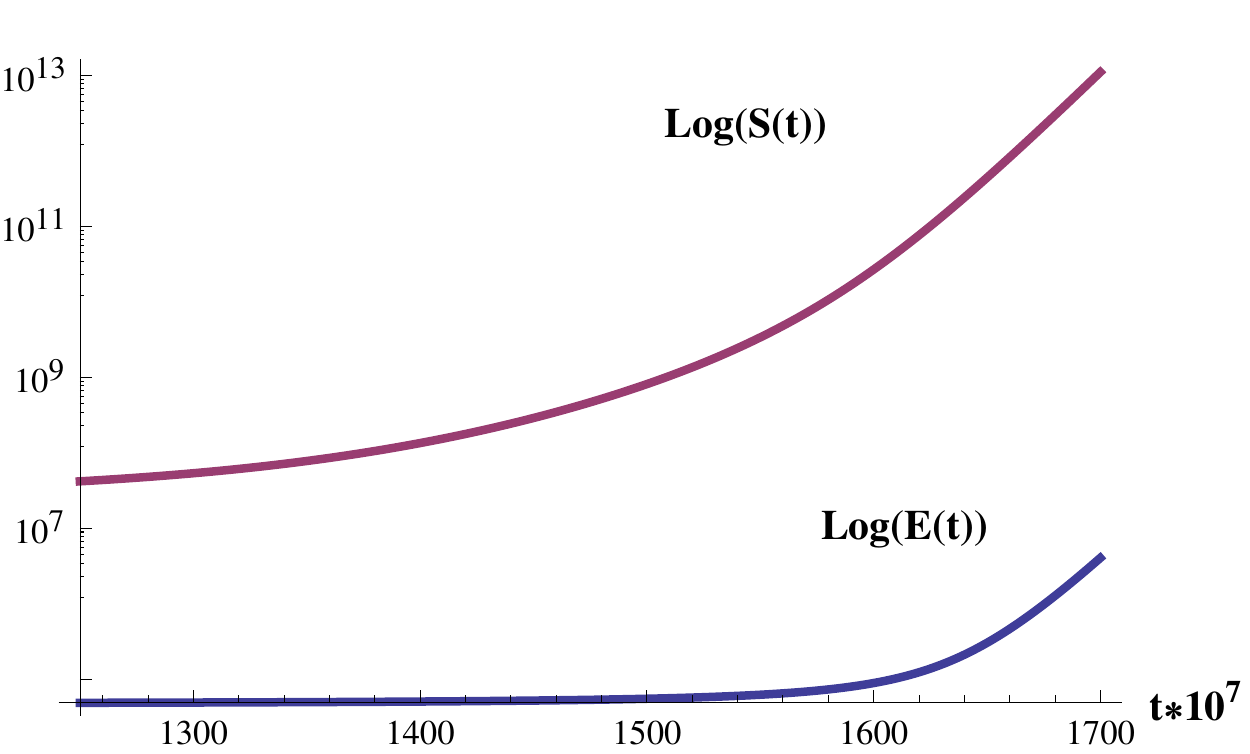}
}  
\caption{\it  Behavior of the total enstrophy $S(t)$ and the total energy $E(t)$. $L=2028$, initial data $\bv_{0}^{+}$.}\label{Fig.11}\end{figure}

 \begin{figure}[H]  
\centerline {
\includegraphics[width=3.5in]{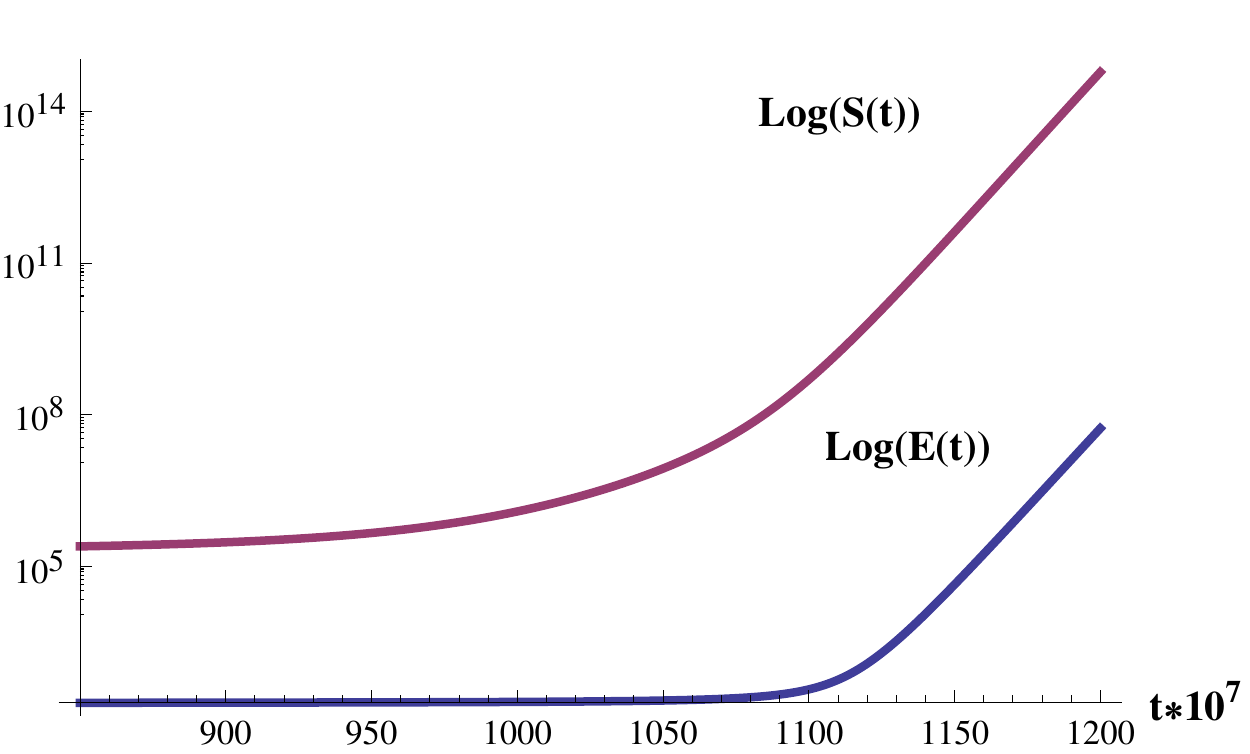}
}  
\caption{\it  Behavior of the total enstrophy $S(t)$ and the total energy $E(t)$. $L=2028$, initial data $\bv_{0}^{-}$.}\label{Fig.22}\end{figure}

 \begin{figure}[H]  
\centerline {
\includegraphics[width=3.5in]{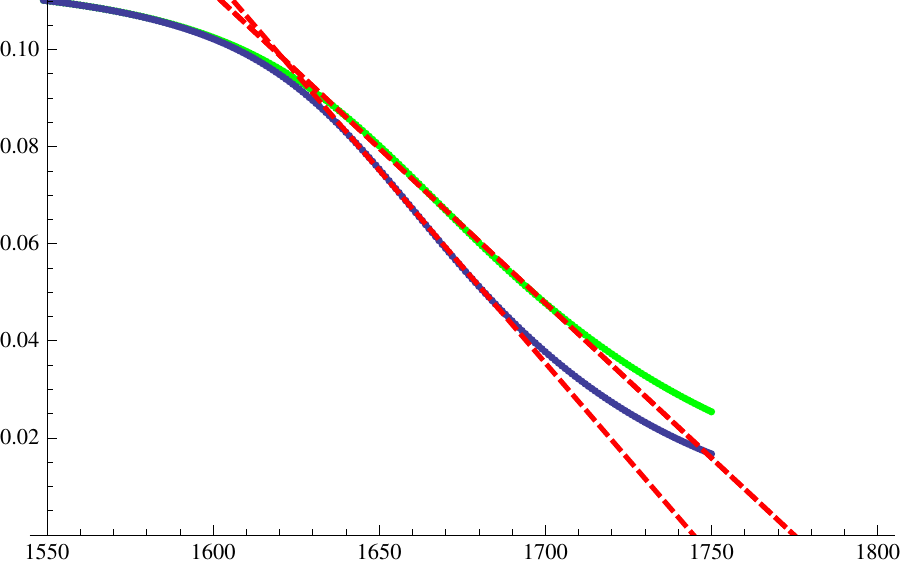}
}  
\caption{\it  Behavior of  $(E(t))^{-{1\over 5}}$ near the blowup for $L=2028$ (green) and $L=2528$ (blue), initial data $\bv_{0}^{+}$. The red dashed line is the linear regression.}\label{Fig.3}\end{figure}

   \begin{figure}[H]  
\centerline {
\includegraphics[width=3.5in]{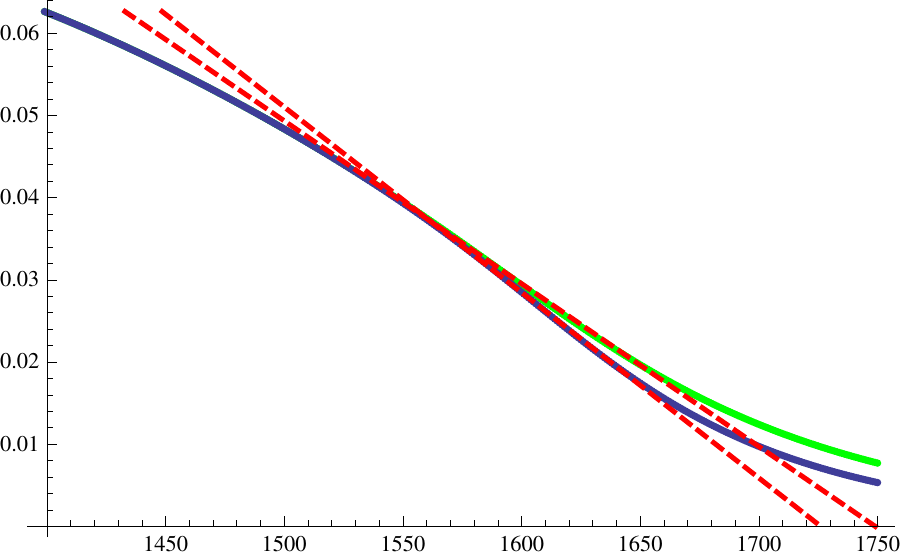}
}  
\caption{\it  Behavior of  $(S(t))^{-{1\over 7}}$ near the blowup  for $L=2028$ (green) and $L=2528$ (blue), initial data $\bv_{0}^{+}$. The red dashed line is the linear regression.}\label{Fig.4}\end{figure}

 \begin{figure}[H]  
\centerline {
\includegraphics[width=3.5in]{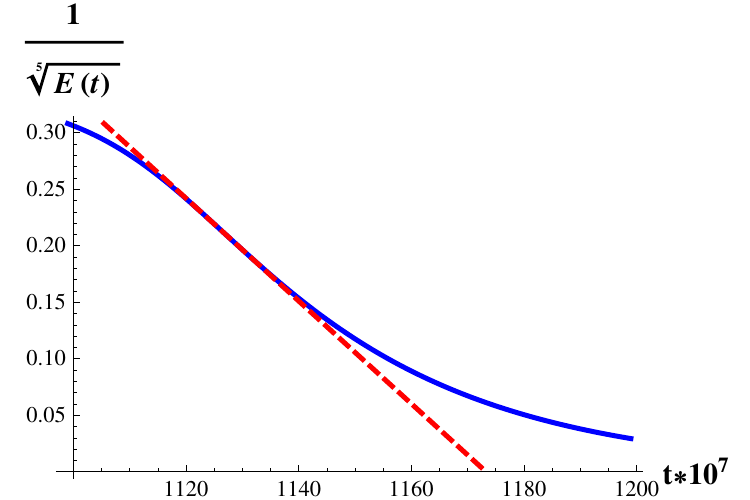}
}  
\caption{\it  Behavior of  $(E(t))^{-{1\over 5}}$ near the blowup for $L=2528$, initial data $\bv_{0}^{-}$. The red dashed line is the linear regression.}\label{Fig.3}\end{figure}

   \begin{figure}[H]  
\centerline {
\includegraphics[width=3.5in]{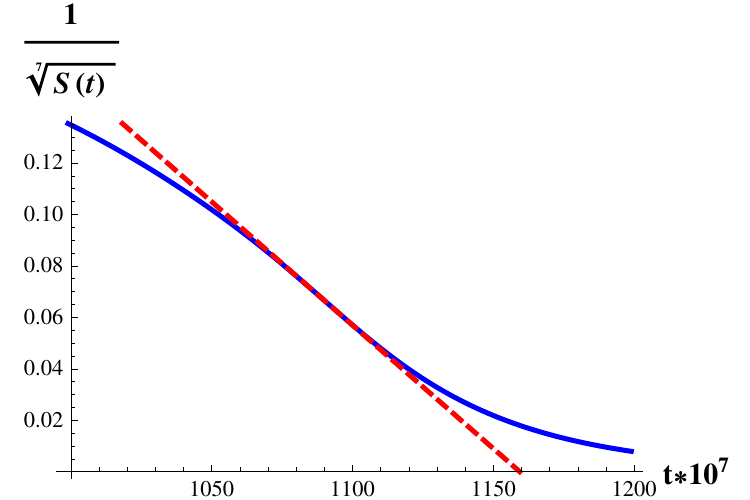}
}  
\caption{\it  Behavior of  $(S(t))^{-{1\over 7}}$ near the blowup  for $L=2528$, initial data $\bv_{0}^{-}$. The red dashed line is the linear regression.}\label{Fig.4}\end{figure}

\section {Behavior of the solution in $\bk$-space.}
\label {S4}

 Figg. 7,8,9 show  the behavior in $\bk$-space of the solutions with initial data $\bv_{0}^{+}$ and $\bv_{0}^{-}$. \par\smallskip
 In Fig. 7 we report the plot of a transversal component of the solution with initial data $\bv_{0}^{+}$, vs $k_{3}$,   for $k_{1}, k_{2}$ and $t$ fixed,  for three values of $t$. It can be seen that the behavior of the solution is close to a damped oscillation with period $2a$, and zeroes at the points $k_{3}\approx {1\over 2} (2j+1)\; a$, $j=1, \ldots$. This is  due to the alternating signs of the series (\ref{8a}).  Moreover the data are compatible with a pointwise convergence in $\bk$ to a limiting function as $t\uparrow \tau$, a fact that is presumably not hard to prove rigorously. \par \smallskip
  A look at Fig. 8, which gives, for the same initial data, the behavior of the marginal density $S_{3}(k_{3},t)$ at $t$ fixed, confirms that the solution has a sequence of zeroes with approximate distance $a$.  On the other hand, for the solution coming out of $\bv_{0}^{-}$, as the coefficients multiplying $\mathbf H^{\zer}({\bk - \bk^{\zer}\over \sqrt p})$ in the series (\ref{8}) are all positive, there are no zeroes and the oscillations are visible only for small $k_{3}$ (Fig. 9). \par\smallskip

  \par\smallskip
 Fig. 10 shows the vectors $\bv(\bk,t)$ on the plane $k_{3}= 100$ at the time $t=1521 \times 10^{-7}$ for initial data $\bv_{0}^{+}$. The picture is limited to the part of the plane were $\bv$ is significantly nonzero. The fact that the velocity field is overwhelmingly radial is a consequence of the fact that the solution corresponds to the fixed point $\mathbf H^{\zer}(\bY)=- 2(Y_{1}, Y_{2}, 0)$. 

 \par\smallskip

The rapid growth of the enstrophy and the time evolution of the main support of the solution along the $k_3$-axis are illustrated by the plot of $S_{3}(k_{3}, t)$ at three different times given in Fig. 11 for $\bv_{0}^{+}$ and in Fig. 12 for $\bv_{0}^{-}$.  
As we mentioned above,  for $t\uparrow \tau$ the support in $\bk$-space moves away to infinity along the $k_{3}$-axis, and gets out of the integration region.  Let $k^{(M)}_{3}(t)$ denote the  value of $k_{3}$ where the maximum of the enstrophy marginal $S_{3}(k_{3},t)$ is located. One can predict, on the basis of the asymptotics (\ref{8}), (\ref{8a}), that, as $t\uparrow \tau$, we should have $k^{(M)}_{3}(t)\sim {C\over \tau-t}$, for some constant $C>0$.  This is shown by Fig. 13, which gives the behavior of $\log [k^{(M)}_{3}(t)(\tau-t)]$ vs $t$ for the initial data $\bv^{+}$, anf  for $\tau = 1640 \times 10^{-7}$. For this choice of the critical time, see the discussion below in this paragraph.  
\par\smallskip

 Concerning the transversal distribution of the enstrophy, Fig. 14 shows the plot of $S_{1}(k_{1}, t) $ for the first two times as in Fig. 11. The transverse marginal $S_{2}(k_{2}, t) $ looks exactly the same. As we see, the support of the solution in the transversal directions grows very slowly and is well contained inside the square $[-127, 127]$ of the computation region. 
  \par\smallskip

   For the initial data $\bv_{0}^{+}$ we will use another  way of predicting the explosion time, based on the decay rate of the maxima of the oscillations (bumps) in the plot of $E(k_{3},t)$, which shows the same kind of oscillations as $S(k_{3},t)$ in Fig.11.   As $\Lambda(t)$ is a smooth function, the expansion (\ref{smooth}) shows that, as $t\uparrow \tau$, for $k_{3}> k^{(M)}_{3}(t)$ the maxima, for $k_{3}$ large enough,    should decay exponentially, with an exponent proportional to $\tau-t$.  \par\smallskip
   Figure 17 shows the logarithmic plot of the maxima of the peaks of $E(k_{3},t)$ versus $k_{3}$ at a fixed time $t$. The points align on straight line with great accuracy, and the slope decreases with time.
Plotting  the slope versus time we obtain a graph, shown on Fig. 18, which is with very   good approximation   a straight line (it almost coincides with its linear regression, indicated by the red dashed line).  The intercept with the horizontal axis, which is around the point $t= 1640 \times 10^{-7}$ appears to be a lower  estimate, presumably more reliable than the upper estimate, of the blow-up time $\tau$.
    
 \par\smallskip
 
 \begin{figure}[H]  
\centerline {
\includegraphics[width=3.5in]{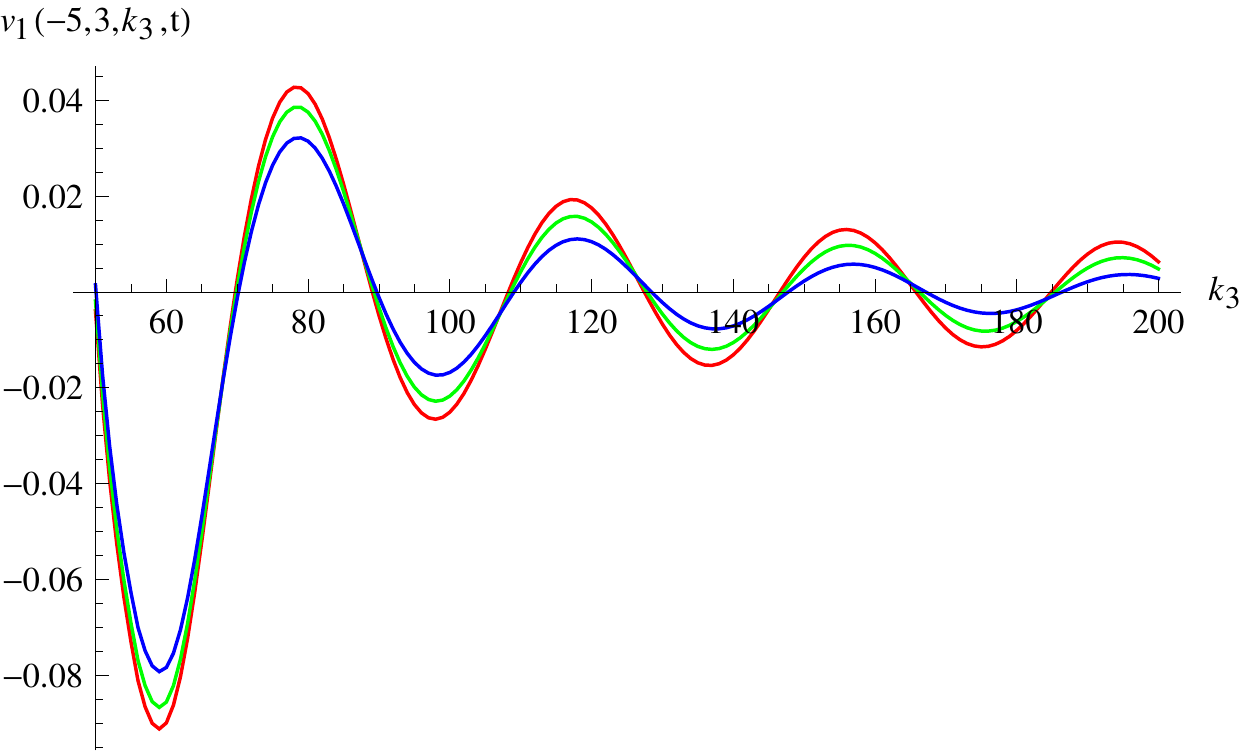}
}  
\caption{\it Plot of $\bv_{1}(\bk, t)$ for $k_{1}, k_{2}$ fixed vs  $k_{3}$, at the times $t=1342 \times 10^{-7}$ (blue),  $t=1500 \times 10^{-7}$ (green), $t=1599 \times 10^{-7}$ (red).  $L = 2028$, initial data $\bv_{0}^{+}$.}\label{Fig.7}\end{figure}

 \begin{figure}[H]  
\centerline {
\includegraphics[width=3.5in]{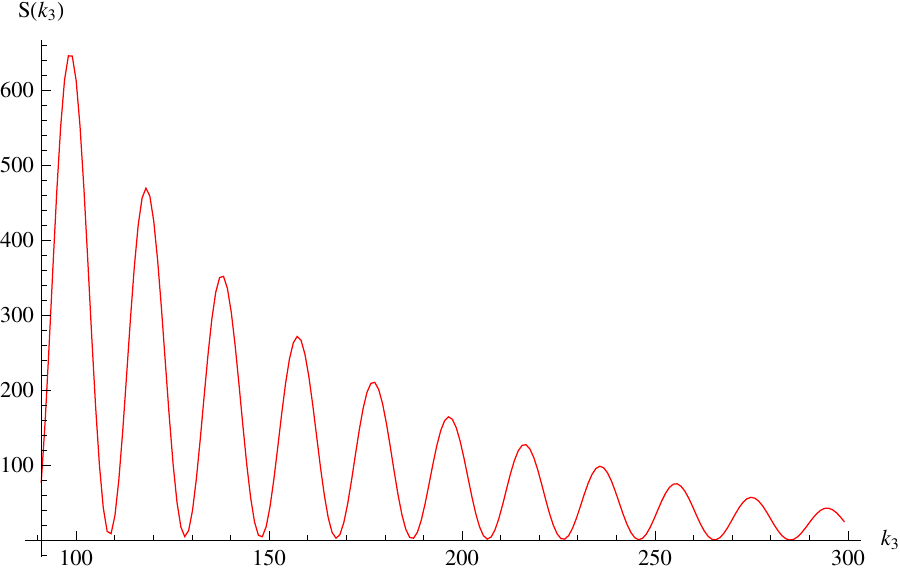}
}  
\caption{\it Plot of $S_{3}(k_{3},t)$ vs $k_{3}$, $t= 1125 \times 10^{-7}$, $L = 2028$, initial data $\bv_{0}^{+}$.} The   zeroes are  approximately periodic with period $a= 20$.  \label{Fig.8}\end{figure}

 \begin{figure}[H]  
\centerline {
\includegraphics[width=3.5in]{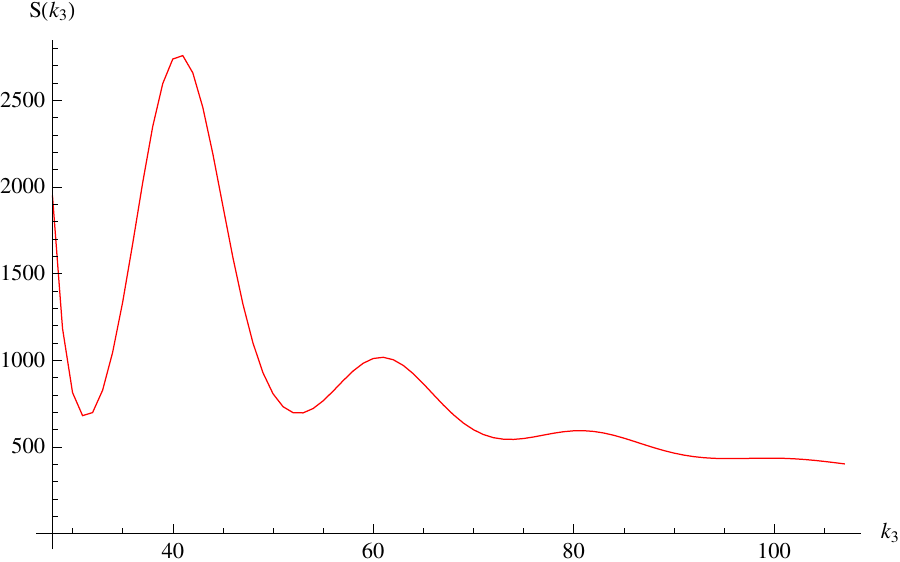}
}  
\caption{\it Plot of $S_{3}(k_{3},t)$ vs $k_{3}$, $t= 900 \times 10^{-7}$.  $L = 2028$, initial data $\bv_{0}^{-}$.}\label{Fig.3}\end{figure}

 \begin{figure}[H]  
\centerline {
\includegraphics[width=3.5in]{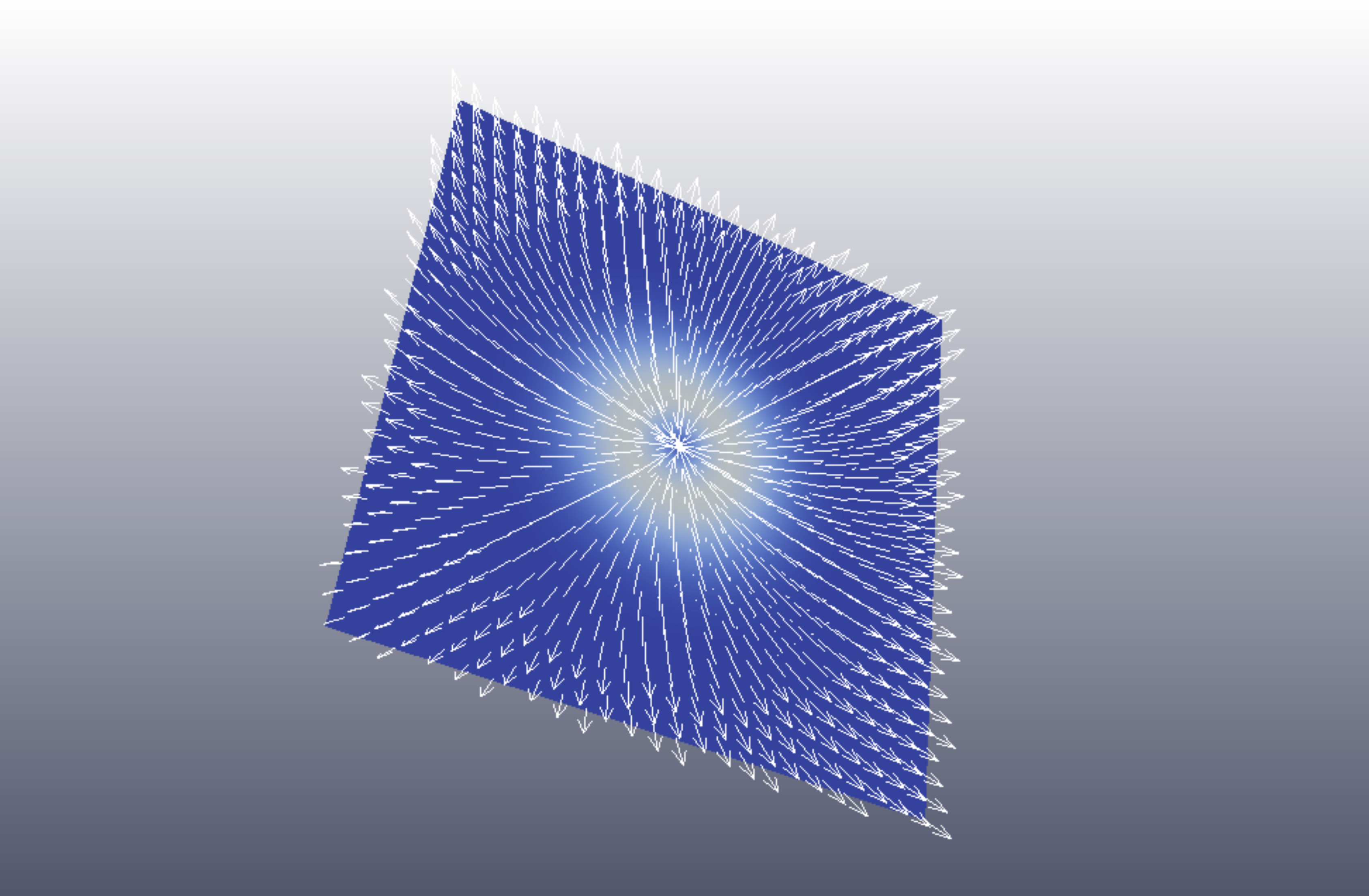}
}  
\caption{\it The vector field $\bv(\bk, t)$ on the plane $k_{3}=100$ and $t=1521 \times 10^{-7}$. The arrows are assigned to a random subset of points and are proportional to the vector norm.  $L = 2528$, initial data $\bv_{0}^{+}$.}\label{Fig.10}\end{figure}

    \begin{figure}[H]  
\centerline {
\includegraphics[width=3.5in]{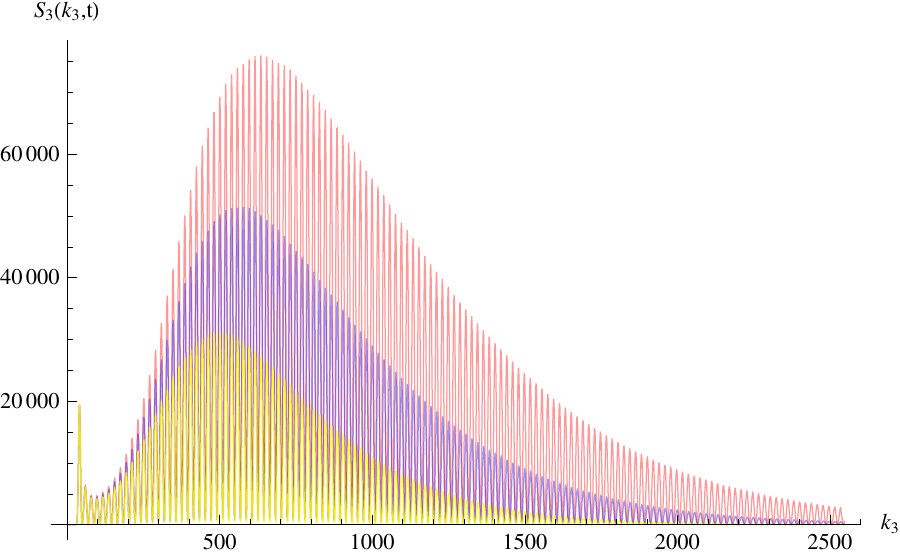}
}  
\caption{\it $S_{3}(k_{3},t)$  at $t \cdot 10^{7} = 1521,  1544, 1560$,  $L= 2528$, initial data $\bv_{0}^{+}$.   }\label{Fig.11}\end{figure}

    \begin{figure}[H]  
\centerline {
\includegraphics[width=3.5in]{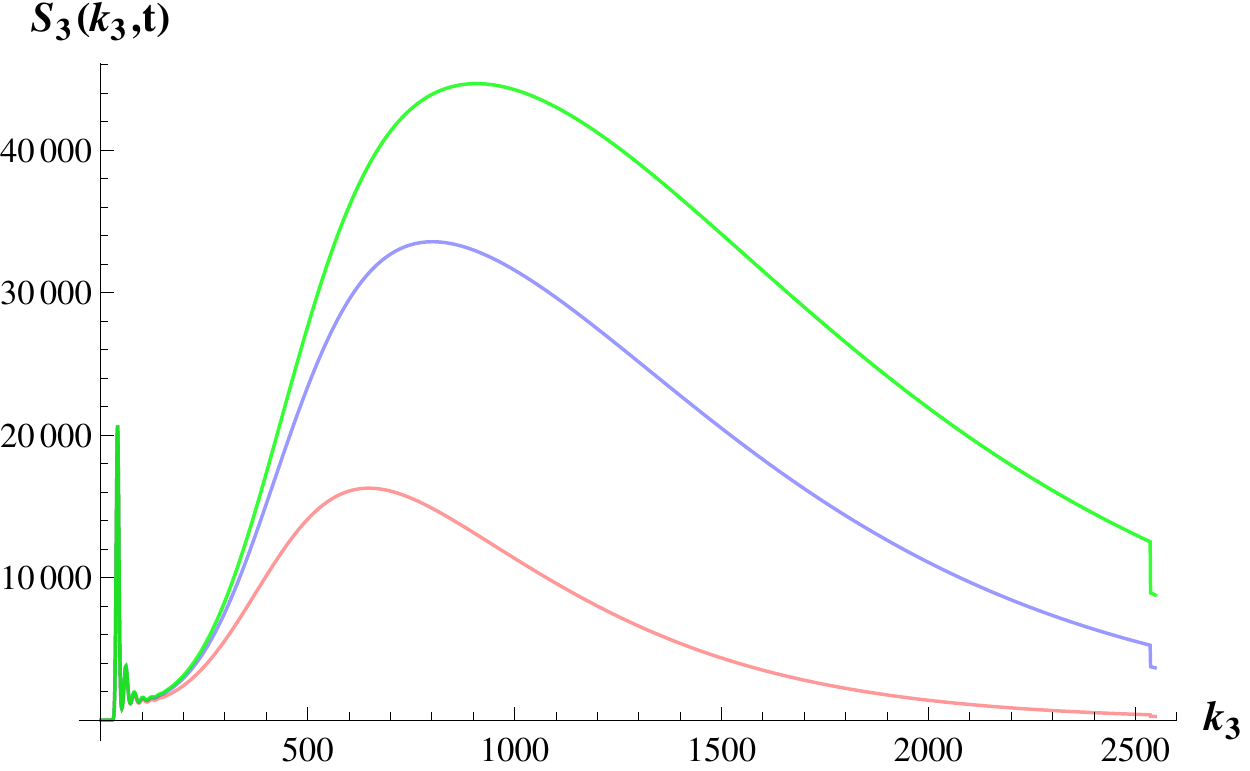}
}  
\caption{\it $S_{3}(k_{3},t)$  at $t \cdot 10^{7} = 1060,  1075, 1080$,  $L= 2528$, initial data $\bv_{0}^{-}$.   }\label{Fig.12}\end{figure}

    \begin{figure}[H]  
\centerline {
\includegraphics[width=3.5in]{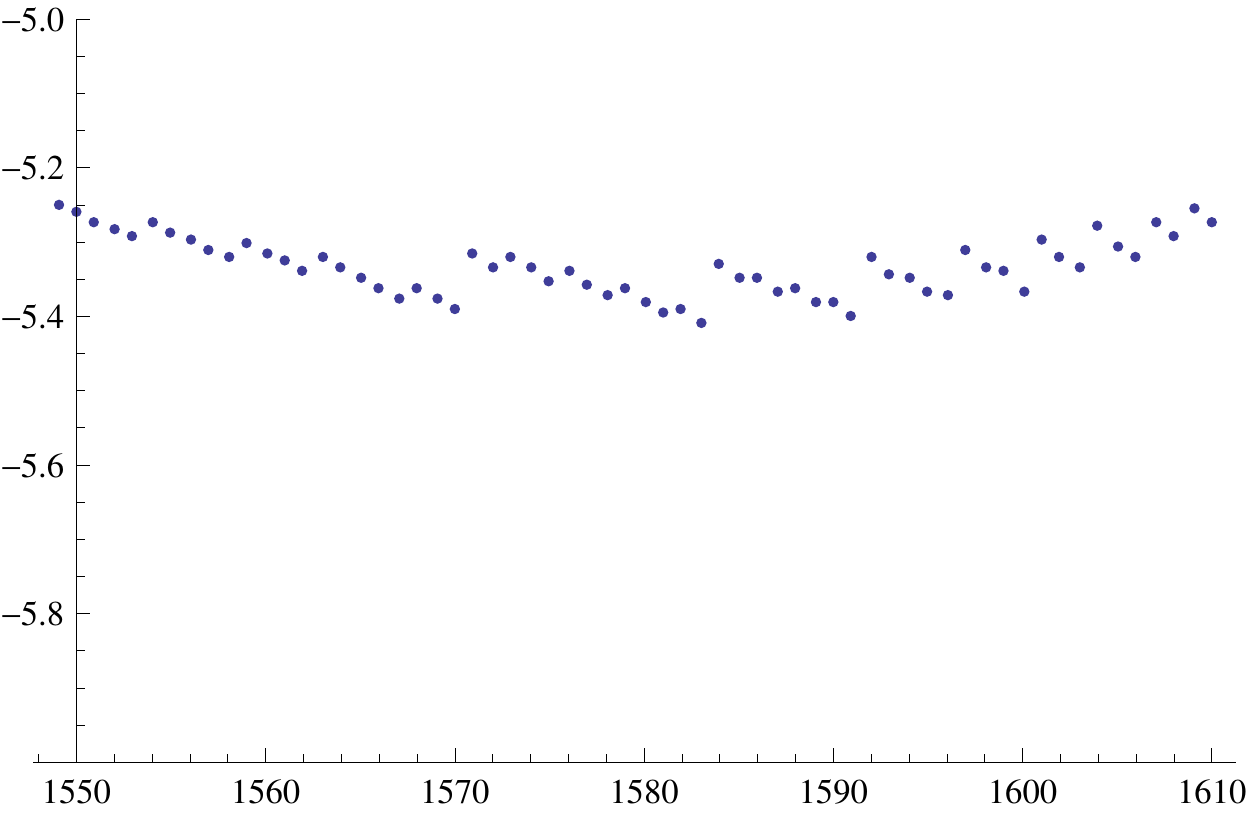}
}  
\caption{\it Behavior of $\log[k^{M}_{3}(t)(\tau-t)]$ vs $t$, $\tau = 1640 \times 10^{-7}$,  $L=2528$, initial data $\bv_{0}^{+}$.}\label{Fig.13}\end{figure}

    \begin{figure}[H]  
\centerline {
\includegraphics[width=3.5in]{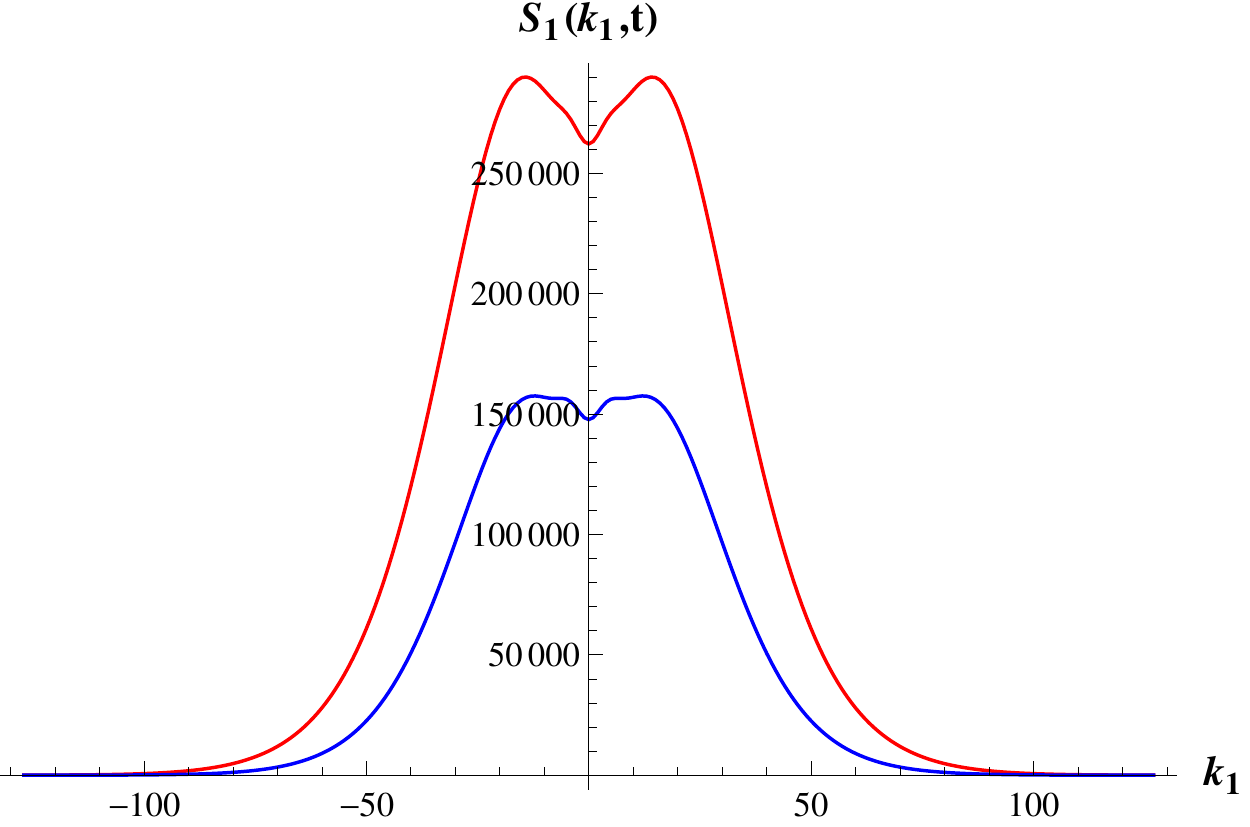}
}  
\caption{\it  $S_{1}(k_{1},t)$  at $t \cdot 10^{7} = 1521,  1544$,  $L=2528$, initial data $\bv_{0}^{+}$.}\label{Fig.14}\end{figure}

 \begin{figure}[H]  
\centerline {
\includegraphics[width=3.5in]{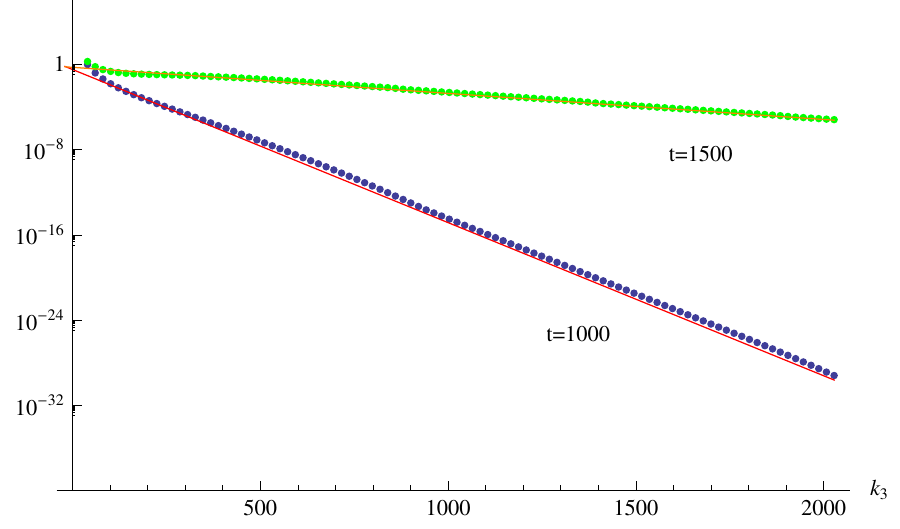}
}  
\caption{\it  Behavior of  the log-plot of the maxima of the oscillations of $E_{3}(k_{3},t)$  before the blowup. $L= 2028$, initial data $\bv_{0}^{+}$.}\label{Fig.15}\end{figure}

 \begin{figure}[H]  
\centerline {
\includegraphics[width=3.5in]{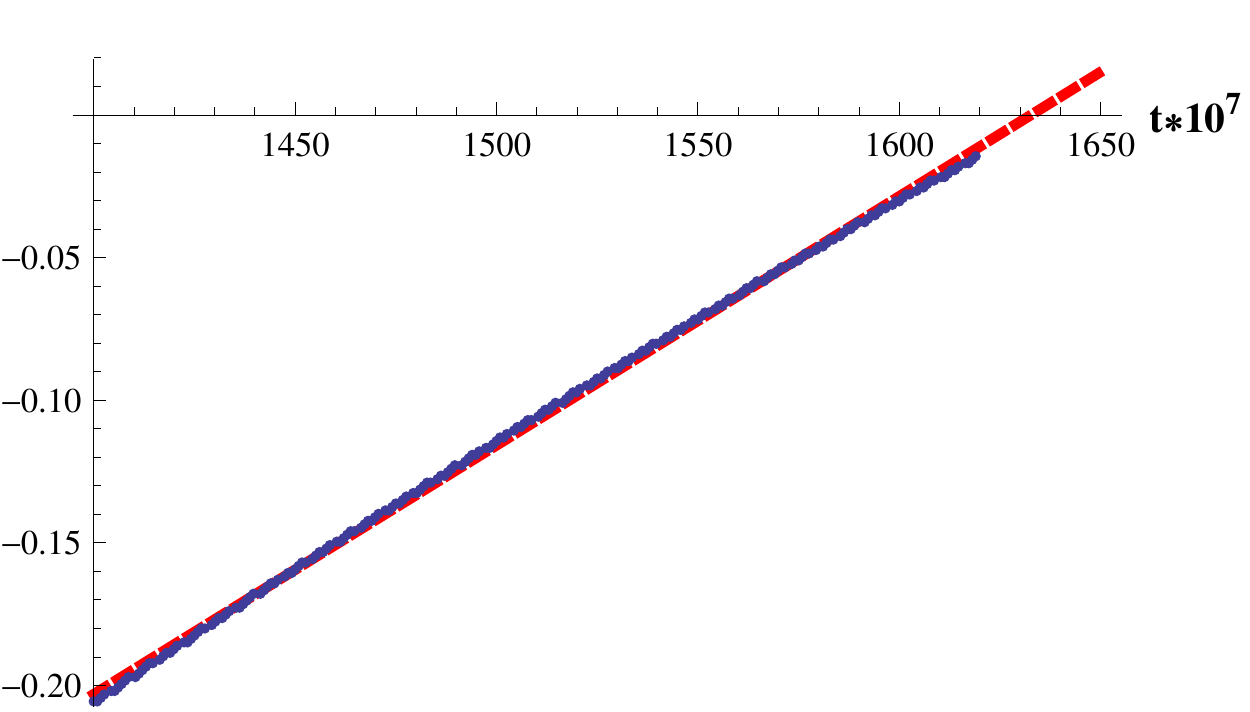}
}  
\caption{\it  Decay rate vs $t\times 10^{7}$ (blue), with  linear regression (red). $L=2528$, initial data $\bv_{0}^{+}$.}\label{Fig.16}\end{figure}

 \section {Behavior of the solution in $\mathbf x$-space}
 \label{S5}
 
The main support  in $\bx$-space of the solution is always contained within a small volume around the origin, and as $t\uparrow \tau$,  it concentrates in sharp ``spikes'' (singular points). The behavior is however qualitatively different for the initial data $\bv_{0}^{+}$ and $\bv_{0}^{-}$.\par\smallskip
 For the initial data $\bv_{0}^{+}$ there are two spikes, at the points 
 $\mathbf x^{\zer}_{\pm}=
 (0, 0, \pm x_{3}^{\zer})$ with $x^{\zer}_{3}\approx 0.16 \approx {\pi\over a}$, as one should expect in view of the fact that, as we saw, $\bv(\bk,t)$ oscilates in $k_{3}$ with a period $T\approx 2a$. Fig. 17 shows the behavior of the energy density $\tilde E_{3}(x_{3},t)$ for $t= 1521 \times 10^{-7}$. \par\smallskip
  At $x= \bx^{\zer}_{\pm} $ the energy grows to $\infty$ as $t\uparrow \tau$, as indicated by Fig. 18, which shows the growth of the maximal values of the marginal distribution  $\tilde E_{3}(\pm x^{\zer}_{3},t)$. Observe that the concentration of energy begins already at $t\approx 1000\times 10^{-7}$, i.e., much earlier than the growth of the total energy.
 \par\smallskip
  
  In Fig. 19 the  plots of the marginal density $\tilde E_{3}(x_{3},t)$ at $t= 1521 \times 10^{-7}$ and $t= 1544 \times 10^{-7}$ are superimposed for comparison. Fig. 20 shows a similar comparison for the marginal density $\tilde E_{1}(x_{1},t)$ at the same times.     \par\smallskip

For the initial data $\bv_{0}^{-}$ we have instead a single spike at the origin, as shown in Fig. 21, which shows the plots of the  marginal $\tilde E_{3}(x_{3},t)$ at two different times. 
 \par\smallskip
 
 The simulation data shown in Figg. 19, 20 and 21  suggest that the marginal densities  $\tilde E_{1}(x_{1}, t)$ and $\tilde E_{3}(x_{3}, t)$ tend to finite limits as $t\uparrow \tau$, except at the singular points.    A possible conclusion, which however would require further investigation, is that   for all $\bx \neq \mathbf x^{\zer}_{\pm}$ the function $\bu(\bx, t)$ itself also converges. Convergence for all points in $\bx$-space except at the singular points was in fact proved for the complex Burgers equations (see \cite{LiSi10} and \cite{BFM12} for computer simulations).

 \par\smallskip
 Except for the fact of showing a single spike, the behavior in $\bx$-space of the solution for initial data $\bv_{0}^{-}$ does not differ significantly, so that in what follows we will only consider the solution with initial data $\bv_{0}^{+}$. 

\par\smallskip
  The marginal distributions of the enstrophy behave in a similar way, except that the spikes are more enhanced than  for the energy marginals. Figg. 22 and Fig. 23 shows the log plot of the marginal $\tilde S_{3}(x_{3},t)$ and  of the transverse marginal $\tilde S_{1}(x_{1},t)$ at a single time.

 \par\smallskip

  An important role   with respect to possible singularities of the NS equations is played by the vorticity stretching  vector
     $\mathbf w(\bx, t) =\omega(\bx, t) \cdot \nabla \mathbf u(\mathbf x, t)$, 
 where $ \omega(\bx,t)  = \nabla \times \bu(\bx,t)$ is the vorticity (see, e.g, \cite{RuG04}).\par\smallskip
 Fig. 24 shows a joint log plot nat the time   $t= 1544 \times 10^{-7}$   of the marginals $\tilde S_{3}(x_{3},t)$ and  
    $$W_{3}(x_{3},t) = \int_{\R\times \R} dx_{1}dx_{2}    \left |  \omega(\bx, t) \cdot \nabla \mathbf u(\mathbf x, t)\right |^{2},$$
    ($W_{3}$, for dimensional homogeneity, is divided by $E_{0}$). It can be seen that $W_{3}$
    is much more concentrated around the two points $\mathbf x^{\zer}_{\pm}$ than the enstrophy marginal $\tilde S_{3}$, which in its turn is more concentrated than the corresponding energy marginal.    
    
 \begin{figure}[H]  
\centerline {
\includegraphics[width=3.5in]{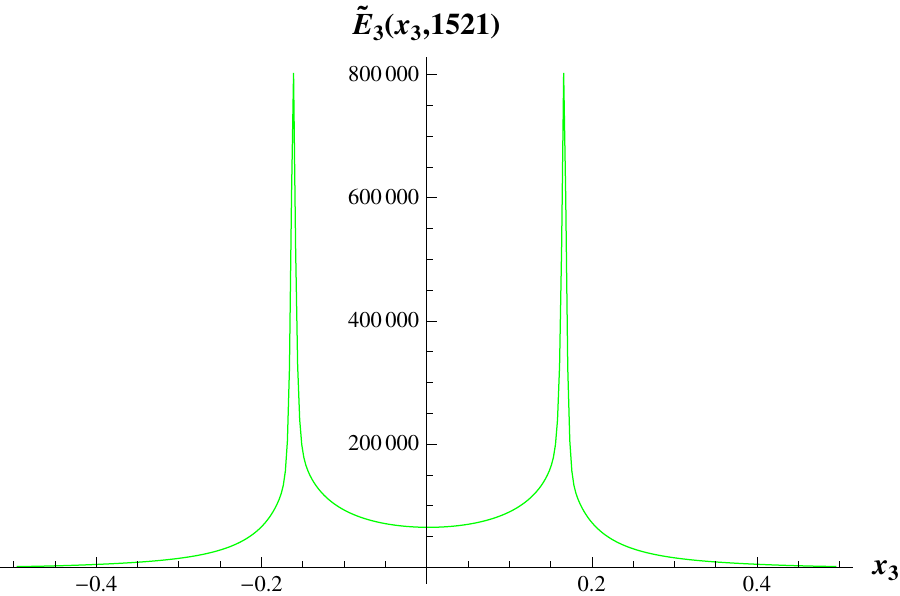}
}  
 \caption{\it  $\tilde E_{3}(x_{3},t)$ at $t \cdot 10^{7} = 1521$. $L=2528$, initial data $\bv_{0}^{+}$.}\label{Fig.17}\end{figure}

 \begin{figure}[H]  
\centerline {
\includegraphics[width=3.5in]{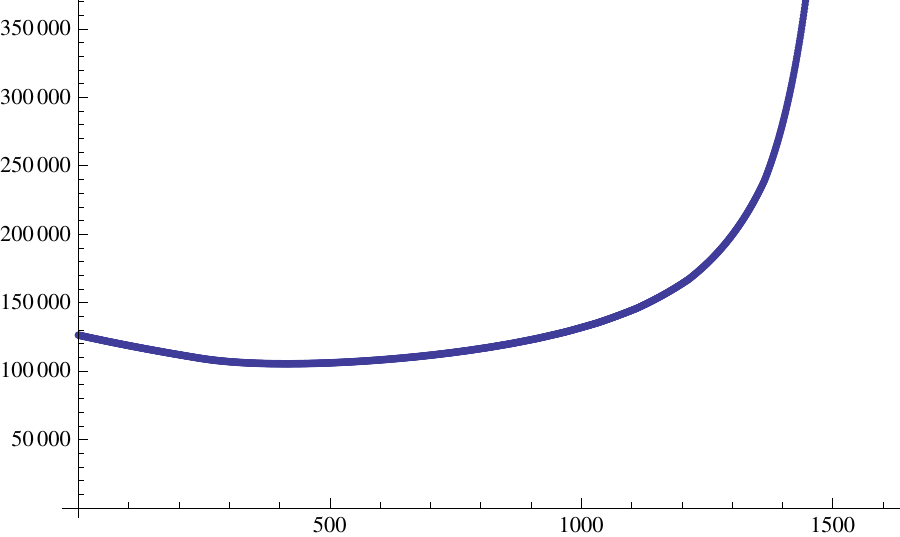}
}  
 \caption{\it  Growth of $\max_{x_{3}}\tilde E_{3}(x_{3},t)$ in time. $L=2528$, initial data $\bv_{0}^{+}$.}\label{Fig.18}\end{figure}

 \begin{figure}[H]  
\centerline {
\includegraphics[width=3.5in]{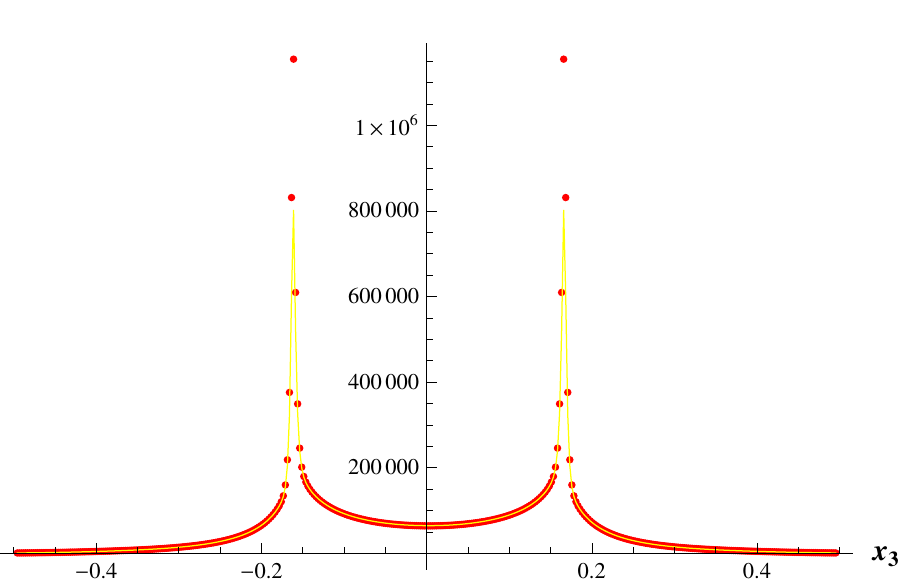}
}  
 \caption{\it  $\tilde E_{3}(x_{3},t)$ at $t \cdot 10^{7} = 1521$ (yellow) and $t\times 10^{7}= 1544$ (red  dots). $L=2528$, initial data $\bv_{0}^{+}$.}\label{Fig.19}\end{figure}
 
 \begin{figure}[H]  
\centerline {
\includegraphics[width=3.5in]{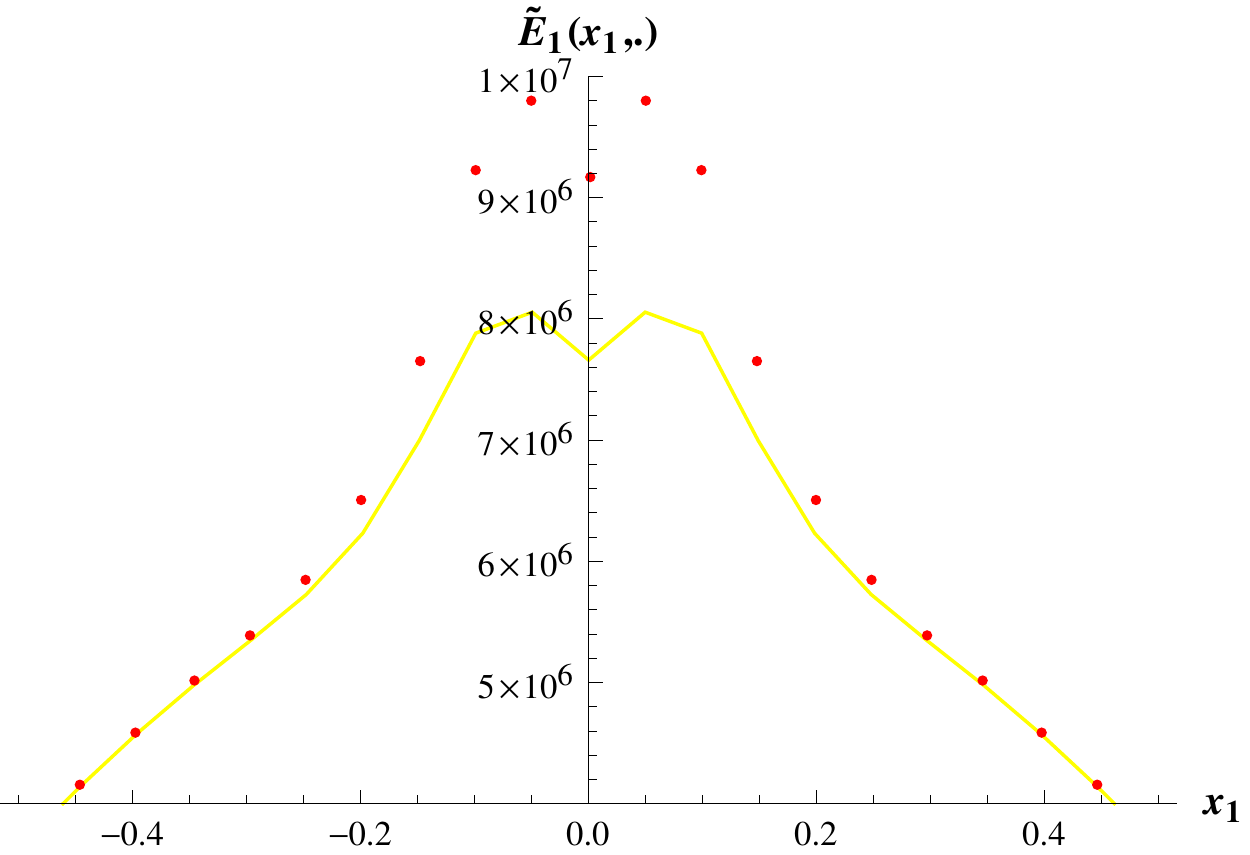}
}  
 \caption{\it  $\tilde E_{1}(x_{1},t)$ at $t \cdot 10^{7} = 1521$ (yellow),  and $t \cdot 10^{7} =1544$ (red dots). $L=2528$, initial data $\bv_{0}^{+}$.}\label{Fig.20} \end{figure}
 
 \begin{figure}[H]  
\centerline {
\includegraphics[width=3.5in]{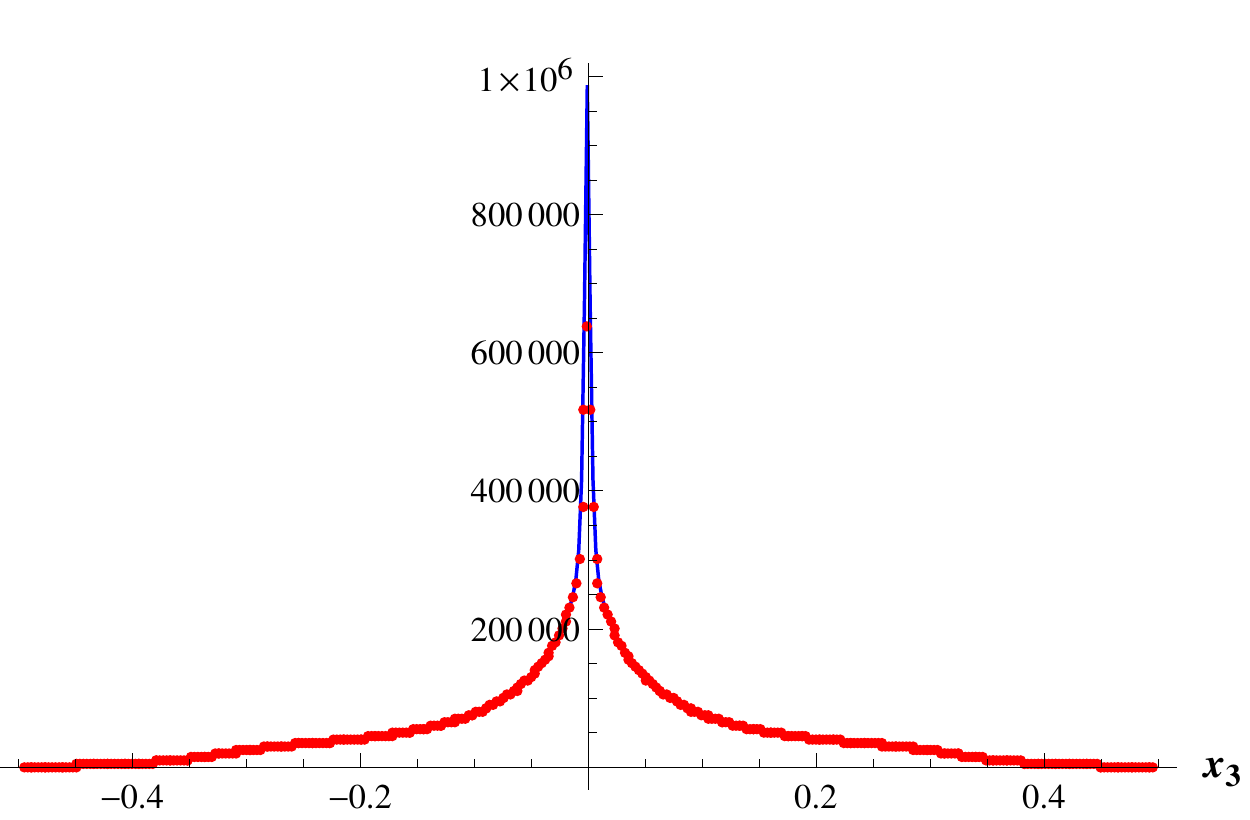}
}  
 \caption{\it  $\tilde E_{3}(x_{3},t)$ at $t \cdot 10^{7} = 1021$ (red dots) and $t\times 10^{7}= 1044$ (blue). $L=2528$, initial data $\bv_{0}^{-}$.}\label{Fig.21}\end{figure}

 \begin{figure}[H]  
\centerline {
\includegraphics[width=3.5in]{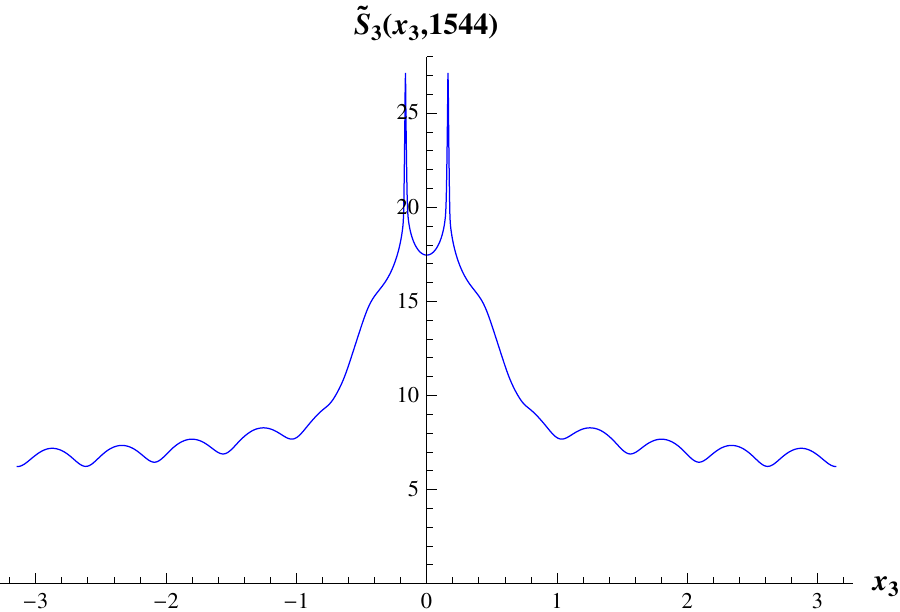}
}  
 \caption{\it Log-plot of $\tilde S_{1}(x_{1},t)$ at $t \cdot 10^{7} = 1544$. $L=2528$, initial data $\bv_{0}^{+}$.}\label{Fig. 22}\end{figure}
 
 \begin{figure}[H]  
\centerline {
\includegraphics[width=3.5in]{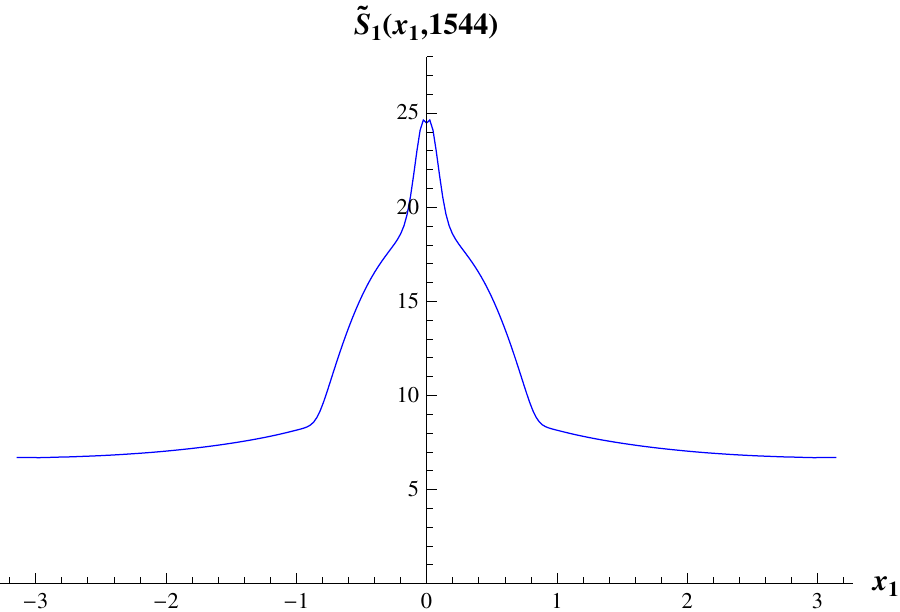}
}  
 \caption{\it Log-plot of $\tilde S_{1}(x_{1},t)$ at $t \cdot 10^{7} = 1544$. $L=2528$, initial data $\bv_{0}^{+}$. }\label{Fig.23}\end{figure}
 
 \begin{figure}[H]  
\centerline {
\includegraphics[width=3.5in]{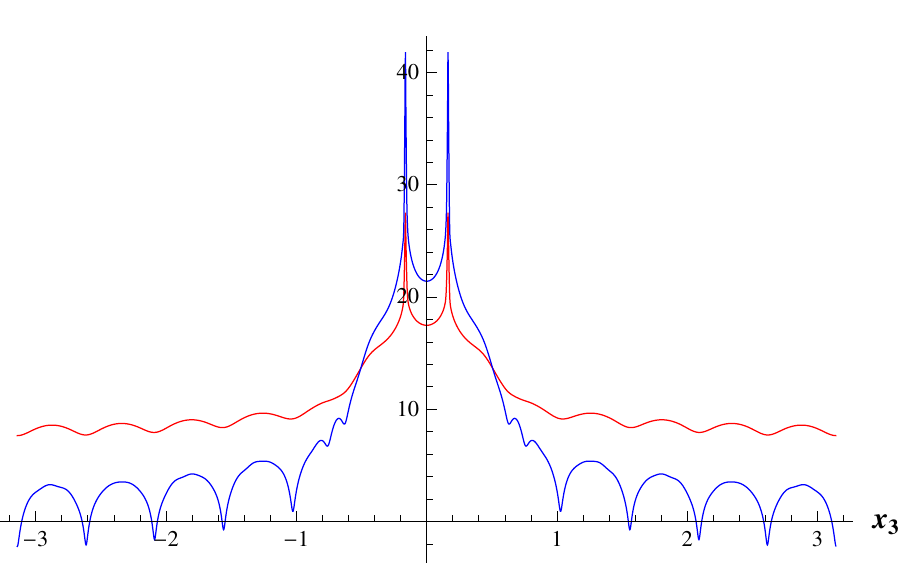}
}  
 \caption{\it Log-plot of $\tilde S_{3}(x_{3},t)$ (red) and $ W_{3}(x_{3},t)$ (blue)  at $t \cdot 10^{7} = 1544$. $L=2528$,  initial data $\bv_{0}^{+}$.}\label{Fig.24}\end{figure}

\section {Some data on the computer simulations·}
\label {S6}

The discretization of our basic equation (\ref{kequation})  starts with a truncation $ [a_{1}, b_{1}] \times [a_{2}, b_{2}] \times [a_{3}, b_{3}]$ of the
integration region for the $\bk$ variables. As we said above, all the results  reported in the present paper are obtained for a uniform step size both in $\bk$ and in  time. \par \smallskip
 The discrete computation
 is obtained by the so-called Nystr\"om method with respect to the
$\bk$ variables and a predictor-corrector scheme, with respect to the time variable,
which is based on the Euler method and the Trapezoidal method.
\par\smallskip
The predictor-corrector scheme iterates a recursive procedure to compute successive approximations $\mathbf V^{(j)}(\bk, n \delta_{t})$, $j=1,2,\ldots$,  of the function $\bv(\bk, n\delta_{t})$, until a convergence criterion is satisfied: $|\mathbf V^{(j+1)}(\bk, n\delta_{t}) -\mathbf V^{(j+1)}(\bk, n\delta_{t}) |\leq {\rm tol}$, where the tolerance ${\rm tol}$ is   set at $10^{-8}$. 
\par\smallskip
The procedure is computationally challenging, and in
fact for each $\bk$ in the mesh it requires the evaluation of a three-dimensional integral. The integral is however
 a convolution,  and by using the fast Fourier transform (FFT) we can  reduce
the computational cost. Note that, as we are simulating the NS equations in $\R^{3}$,  the convolution is not a periodic convolution and has to be
 implemented on a computational grid which is doubled in size.
\par\smallskip

The accuracy of the approximated solution corresponding to the chosen initial data is evaluated on an experimental basis by
comparing the results obtained for different discretization parameters.

\par\smallskip

  Our computer simulations were performed at
   CINECA of Bologna (Italy) on  the  FERMI Supercomputer (Model: IBM-BlueGene/Q; Architecture: $10$ BGQ Frame; Processor
Type: IBM PowerA2, $1.6$ GHz; Computing Cores $163840$; Computing Nodes
$10240$; RAM: 1GByt/core)   
   \par\smallskip
The computation method was implemented in Fortran 90 (IBM Fortran compiler)
with MPI library for parallel computations, and 2Decomp\&FFT for the
parallel computation of the fast Fourier transform.

 \section {Concluding Remarks}
 \label{S7}
 \par\smallskip
 
 Simulations of the solutions of the 3-d NS equations are usually computationally onerous and sometimes unreliable, especially for flows with large values of the enstrophy and of the vorticity stretching.     It is a remarkable fact that the singular complex  solutions proposed by Li and Sinai in \cite {LiSi08}, due to their simple structure in $\bk$-space, are relatively easy to follow  by computer simulations on the supercomputers of the last generation.  
 
 \par\smallskip
 Of particular importance in this respect is the representation of the solution as a power series (\ref{serie}), where the parameter $A$ governs the blow-up time. A great help also comes from the stability of the computation with respect to the discretization step in $\bk$-space, which is due to the confinement of the energy in a small region of $\bx$-space, as we discussu below.  It is perhaps worth to observe that in the paper of Li and Sinai  \cite{LiSi08} the infinite extension of the domain in $\bx$-space does not seem to be  essential, except for the fact that there are no boundary conditions, and that the proofs can be adapted,  by some straightforward modifications, to the periodic case on the torus $T^{3}$ as well.
 
  \par\smallskip

 The results of  our computer simulations give a clear evidence of the blow-up, with an estimate the critical time. It was also possible to obtain a detailed picture of the behavior of the solutions near the critical time,  confirming the predictions of the theory, and  also producing evidence of important properties, such as the pointwise  convergence as $t\uparrow \tau$  of the solution in $\bk$-space, and  in $\bx$-space except for the singular points. Whether such properties really hold  requires further study.
\par\smallskip
  The general picture that comes out is that of a   motion in which the fluid points move very fast in a small region around the origin, for the initial data $\bv_{0}^{-}$, or around
   two symmetric points close to the origin, for the initial data $\bv_{0}^{+}$,  along flow lines with high curvature. Fig. 25 shows the flow lines (for the real part of $\bu(\bx,t)$) starting at $t=0$ from a random set of fluid points in a small central region and up to time 1521. One can see that  the overwhelming majority of the lines  is confined to the central region.
 \par \smallskip
 
 As we go to the critical time, the enstrophy, and even more the vorticity stretching, are increasingly in spikes, which diverge to infinity at the singular points, while the fluid at a finite distance remains ``quiet.'' A behavior which reminds that of a physical tornado.\par\smallskip

 The results so far obtained suggest the study of real-valued solutions of the 3-d NS equations which share some basic properties, such as the extension of the essential support of the solutions in $\bk$-space, with the complex solutions considered in the present paper. More work in this direction is in progress. 
\par\vskip 1.0 cm
 
 \begin{figure}[H]  
\centerline {
\includegraphics[width=3.5in]{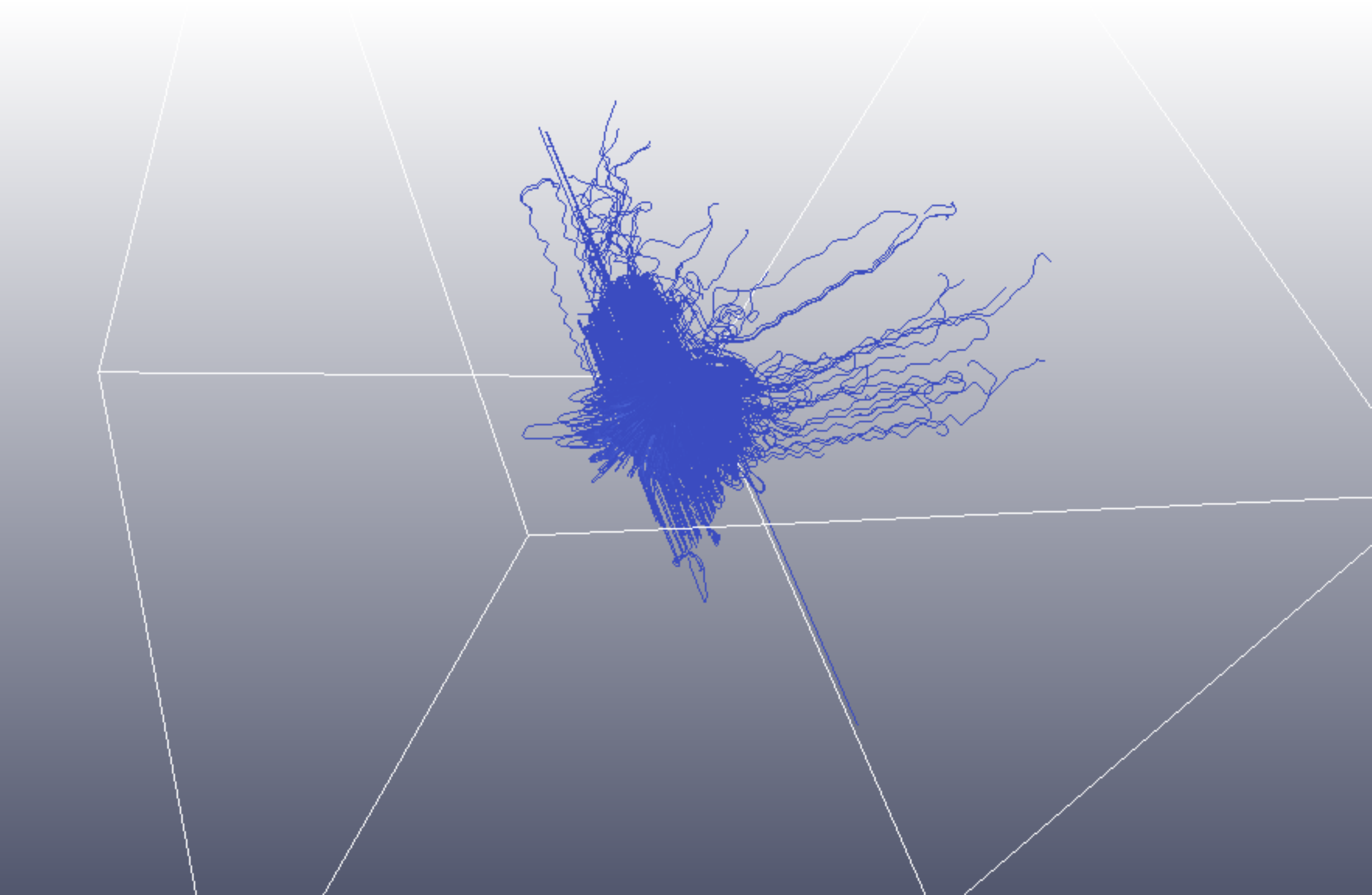}
}  
 \caption{\it Flow lines from a random set of points up to time $1541 \times 10^{-7}$. $L=2528$, initial data $\bv_{0}^{+}$. }\label{Fig.25}\end{figure}

\section {Acknowledgements} 
\label{S8}
We thank Prof. Ya. G. Sinai for his constant interest on our work, and for many discussions and suggestions. We also thank Dr D. Li for valuable remarks. We acknowledge the CINECA award under the ISCRA initiative IsB10\_3DNS (2014) and IsC23\_3DNS (2015),  for the availability of high performance computing resources and support.
    
   \vfill\eject


\begin{thebibliography}{99}
    
    \bibitem{AKh09}  Arnol'd, M.D. and Khokhlov, A.V.: ``Modeling a blow-up solution of tornado type for the complex version of the three dimensional Navier-Stokes equation''. Russian Mathematical Surveys, {\bf 64}, 1133--1135, 2009\par
  
    
  \bibitem{BFM12} Boldrighini, C., Frigio, S. and Maponi, P.: ``Exploding solutions of the two-dimensional Burgers equations: Computer simulations''.  J. Math. Phys. {\bf 53},083101, 2012\par
  
    \bibitem{Ch08} Cheskidov, A.:  ``Blow-up in finite time for the dyadic model of the Navier-Stokes equations''. Trans. Am. Math. Soc., 10, 5101-5120, 2008\par

\bibitem{Hou09} Hou, Th. Y.: ``Blow-up or no blow-up? A unified computational and analytic approach to  three-dimensional incompressible Euler and  Navier-Stokes equations''.  Acta Numerica, {\bf 18}, 277-346, 2008 \par
 \bibitem{KP02} Katz, N.  Pavlovic, N.: ``A cheap Caffarelli-Kohn-Nirenberg inequality for the Navier-Stokes equation
with hyper-dissipation''.  Geom. Funct. Anal. {\bf 12}, No. 2, 355-379, 2002
\par

    \bibitem {Leray} Leray, J.: ``Sur le mouvement d'un liquide visqueux emplissant l'\'espace''.   Acta Math   {\bf 63}, 193--248, 1934\par
    \bibitem{Temam} Temam, R.: {\it Navier-Stokes Equations.} North Holland,  1979\par

   \bibitem{LiSi08}   Li, D. and Sinai, Ya. G.:    ``Blowups of complex solutions of the 3D Navier-Stokes system and renormalization group method''.    J. Eur. Math. Soc.  {\bf 10}, 267--313, 2008 \par 
 \bibitem{LiSi10} Li, D, and Sinai, Ya.G.:  ``Singularities of complex-valued solutions of the two-dimensional Burgers system''. J. Math. Phys. {\bf 51}, 01525, 2010 \par
  \bibitem{LiSi10(2)} Li, D, and Sinai, Ya.G.: ``Blowups of Complex-valued Solutions for Some Hydrodynamic models''. Regular and Chaotic Dynamics {\bf 15}, Nos 4-5, 521-531, 2010 \par
  
  \bibitem {RuG04} Ruzmaikina, A, and Grujic, Z.: ``On Depletion of the Vortex-Stretching Term
in the 3D Navier-Stokes Equations''. Comm. Math. Phys, {\bf 247} , 601-611,  2004
 \par
 \bibitem{Seregin12} Seregin, G.: ``A Certain Necessary Condition of Potential Blow up for Navier-Stokes Equations''. Commun. Math. Phys. {\bf 312}, 833-845, 2012
 \par
 
 
  \bibitem{TT14} Tao, T.: ``Finite time blowup for an averaged three-dimensional Navier-Stokes equation''.  arXiv: 1402.0290v2 [math AP], 2014\par
  
 



 

 
 \end{thebibliography}
       \end{document}